\documentclass[%
 reprint,
superscriptaddress,
showkeys,
 aps,
 pra,
]{revtex4-1}

\usepackage[margin=1in]{geometry} 
\usepackage{amsmath,amsthm,amssymb}
\usepackage{graphicx}
\usepackage{braket}
\usepackage{dsfont}
\usepackage{comment}
\usepackage{blkarray}
\usepackage{algorithm}
\usepackage[noend]{algpseudocode}
\usepackage{graphicx}
\usepackage{dcolumn}
\usepackage{bm}
\usepackage{amsfonts}
\usepackage{epsf}  
\usepackage{appendix}
\usepackage{mathtools}

\usepackage{hyperref}
\usepackage[mathlines]{lineno}

\newtheorem{Problem}{Problem}
\newtheorem{Lemma}{Lemma}
\newtheorem{Theorem}{Theorem}

\newtheorem{Definition}{Definition}
\makeatletter
\def\BState{\State\hskip-\ALG@thistlm}

\usepackage{xcolor}

\begin{document}

\preprint{APS/123-QED}

\title{A quantum active learning algorithm for sampling against adversarial attacks}

\author{P. A. M. Casares}
 \email{pabloamo@ucm.es}
 \affiliation{Departamento de F\'isica Te\'orica, Universidad Complutense de Madrid.}
\author{M. A. Martin-Delgado}%
 \email{mardel@ucm.es}
\affiliation{Departamento de F\'isica Te\'orica, Universidad Complutense de Madrid.}%



\date{\today}

\begin{abstract}
Adversarial attacks represent a serious menace for learning algorithms and may compromise the security of future autonomous systems. A theorem by Khoury and Hadfield-Menell (KH), provides sufficient conditions to guarantee the robustness of machine learning algorithms, but comes with a caveat: it is crucial to know the smallest distance among the classes of the corresponding classification problem. We propose a theoretical framework that allows us to think of active learning as sampling the most promising new points to be classified, so that the minimum distance between classes can be found and the theorem KH used. Additionally, we introduce a quantum active learning algorithm that makes use of such framework and whose complexity is polylogarithmic in the dimension of the space, $m$, and the size of the initial training data $n$, provided the use of qRAMs; and polynomial in the precision, achieving an exponential speedup over the equivalent classical algorithm in $n$ and $m$. This algorithm may be nevertheless `dequantized' reducing the advantage to polynomial.\footnote{In the original article published in New Journal of Physics it is mentioned that our article was likely dequantizable. Here we have added a small appendix to explain how.}

\end{abstract}

\pacs{Valid PACS appear here}
\maketitle


\section{\label{sec:intro}Introduction}

Supervised learning is one of the subareas of machine learning \cite{inner_product,schuld2015introduction,biamonte2017quantum} that consists of techniques to learn to classify new data taking as example a training set. More specifically, the computer is given a training set $X$, consisting on $n$ pairs of point and label, $(x,y)$. With the information, the computer is supposed to extract or infer the conditional probability distributions $p(y|x)$ and use it to classify new points $x$. This paradigm is in contrast with unsupervised learning, that like in the case of clustering, attempts to find structure to a set of points without labels; and reinforcement learning \cite{paparo2014quantum}, where an agent has to figure out the best policy or action for each situation it may face.

An important kind of supervised learning is what is usually called \textit{active learning} \cite{melnikov2018active}. To introduce this concept suppose that we have a supervised learning algorithm, with its corresponding training set. However, instead of directly trying to predict the label of new points, we give the classifier the option to pose us interesting questions in order to reduce the uncertainty in $p(y|x)$. In this setting, the algorithm will add to its training set new points in areas where it has a lot of uncertainty. 

To explain the concept better, let us give an example. Suppose we have an image classifier used for a self-driving car, that has to to distinguish between cars, pedestrians... Internally, images are decomposed into pixels that can be characterised by their combination of red, blue and green. Thus, an image can be expressed as an array of 3 dimensional vectors, or as a large vector if we flatten the array. Any image is then a vector which can be identified with a point of a high dimensional space containing all images.
The set of images of different classes forms a single manifold $\mathcal{M}$ embedded in that highly dimensional space. In particular, the dimension of the space is $3n_p$, for $n_p$ the number of pixels. The key idea of an active learning algorithm is one that is able to understand in what kind of images it has the most uncertainty, and request additional examples to be labeled and added to its training set. 

Another important concept in the context of supervised learning is that of \textit{adversarial attacks} or \textit{adversarial examples}, the name given to a phenomenon where a trained and accurate (usually a neural network) classifier, can be mislead into wrong classification, by producing a carefully chosen and slightly modified version of one point that is classified well. Adversarial examples were discovered quite recently \cite{Intriguing_properties,goodfellow2014explaining}, and have received a lot of attention leading to models robust to particular attacks \cite{Madry2018,sinha2017certifying,wong2017provable,raghunathan2018certified}. 

After the discovery of such adversarial examples, considerable effort has been put in explaining why they happen and how they can be avoided. One of the given theoretical reasons links their existence to a high codimension, the difference in dimension in the structure of the classes we are trying to separate with respect to the highly dimensional space in which they are embedded \cite{Geometry_of_adversarial_examples}. Intuitively this means that the dimension of the space of images of cars, for example, is much lower than the entire space of all possible images, of dimension $3n_p$ as said earlier; since we need to impose a lot of constraints for an image to be indeed the image of a car, even if such constraints are not easily definable. 

This suggest a strategy, proposed in theorem 5 of [12], that states that if we are able to cover our classes with a sufficiently fine sampling of the classes, our algorithm will be provably robust against these adversarial examples. However, how fine this sampling is depends crucially on the minimum distance $r_p$ between classes, and since we do not fully know the classes, we also do not know this minimum distance with precision. Overestimating $r_p$ will result in not being able to use theorem 5 of [12]; whereas underestimating it will mean oversampling the classes, with the associated cost.

In this article we present a quantum active learning algorithm that allows for fast sampling of the most informative points that could be added to the training set $X_\mathcal{L}$ to find out this minimum distance $r_p$. Here, the concept of `informativeness' will refer to an expected gain of information, in the sense of improving the estimate of $r_p$. Our aim is to sample points to be added to the training set with the highest possible informativeness. It will be defined as the product of the probability that a given point is in a class, and the inverse of the margin of the Quantum Support Vector Machine, which measures the amount of information gained if the point were really in that class.

The quantum algorithm will allow us to perform this sampling very efficiently, in polylogarithmic time in the dimension of the space $m$ and the number of already classified points, $n$, and polynomial cost in all other variables. To achieve such complexity we will need to use qRAMs \cite{QRAM_bucket_brigade} and techniques from \cite{QSVM} and \cite{QLSAchilds}. By comparison, linear algebra operations will require polynomial cost in $n$ and $m$, when performed using classical computing.

The origin of this advantage is that in order to calculate such `informativeness' we need to solve the Quantum Support Vector Machine, but we will not need to read out the solution. Rather, we will operate quantumly with the output to calculate the scalar value of `informativeness', which we define formally in the next section. Notice that most quantum linear algebra algorithms are better suited for the cases where one does not have to read out the solution entirely, but rather calculate some expected value. This is partially the intuition that motivated our research. 

We also hope that the framing of this article will highlight to the quantum machine learning community, often focused mostly on obtaining quantum advantages, the necessity of designing systems that are not only efficient and capable, but also robust and reliable.

The structure of the remaining sections of the article is the following. In the following subsection \ref{sec:related} we introduce related work on the topics of adversarial examples, active learning, and quantum machine learning; we also briefly mention the differences between adversarial examples and generative adversarial networks. In section \ref{sec:active_learning} we introduce a bit more of background on theorem 5 of \cite{Geometry_of_adversarial_examples}, and explain how to model the problem of finding the minimum distance between classes as an active learning problem. In particular we introduce the important concept of `informativeness'. Finally, some background on Support Vector Machines is reviewed in \ref{sec:background_SVM}. 

 Section \ref{sec:Main algorithm} constitutes the main corpus of the text. To perform the active learning algorithm, in \ref{sec:Probability} we explain how to obtain $P_c(\vec{x})$, the probability that an arbitrary point of the space is in a given class, which is one of the main components of the `informativeness' of such point. The other main component is described in the next two subsections. In section \ref{sec:QSVM} we review the main results that we will be using from our reference \cite{QSVM}, and in \ref{sec:norm} we use the result of the previous section to calculate $|\vec{w}|_{n+1}$, the second main component of the `informativeness' of $\vec{x}$. Subsection \ref{sec:Target} explains the strategy to select a point that with high probability improves the current estimate of the SVM. Section \ref{sec:results} reviews the main results, and section \ref{sec:Complexity} is dedicated to the calculation of the complexity of our algorithm. Finally, in section \ref{sec:Conclusion} we explain our conclusions. In the appendices we include definitions, theorems from other articles that we use, and some technical results.

\subsection{\label{sec:related}Related work.}

Adversarial examples are a danger for any classifier that needs to be robust to perturbation. For instance, adversarial examples can be dangerous when an autonomous car has to recognise traffic signals.
Since adversarial examples were discovered \cite{Intriguing_properties}, there has been lots of work to explain why they happen \cite{goodfellow2014explaining} and also to obtain provably robust models \cite{Madry2018}. In particular some of the most promising ideas to avoid them are related to adversarial training: training against those adversarial examples before the actual adversary has time to pose them to the classifier. This is for instance the model explained in \cite{Madry2018}, where they use projected gradient descent to minimize the maximum expected loss
\begin{equation}
    \min_\theta \left(\mathbb{E}_{(x,y) \sim \mathcal{D}} \left[\max_{\epsilon \in \mathcal{S}} L(\theta,x+\epsilon,y)\right]\right),
    \label{madry minimization}
\end{equation}
where $\mathcal{D}$ is the initial population of points $x$, with true label $y$, and perturbations $\epsilon$ can be taken from a small set $\mathcal{S}$. $L$ is the loss function: the function that measures the difference between the predicted and actual classification, with adjustable parameters $\theta$, and $\mathbb{E}$ indicates expected value. The maximisation represents the work of the adversary, whereas the minimization represents the work to make the classifier robust. This setup certainly works, as long as one can minimize \eqref{madry minimization}, and efforts have been put forward to perform this adversarial training more efficiently \cite{qin2019adversarial}. However, as pointed out in \cite{Geometry_of_adversarial_examples}, in order to work perfectly it would require an exponential number of adversarial examples in the dimension of the problem added to the training set. Thus, additional strategies are worth exploring.  

It is also worth noticing that adversarial attacks are gaining attention in the quantum community. Recently, it has been indicated that this phenomenon is also present in the case of quantum classifiers \cite{quantum_adversarial}, where the dimension plays a very important role: the higher the dimension the easier to carry out those adversarial attacks. Some experimental work on this line is done in \cite{lu2019quantum}.

Our work is additionally strongly related to several forms of \textit{active learning} algorithms. Active learning algorithms are those where the classifier can ask for new points to be classified and added to its training data base. This field can be divided in two main branches\cite{wang2015active}: \textit{query synthesis}, where new examples are created, and \textit{sampling}. The later is subdivided in \textit{stream-based sampling} and \textit{pool-based}. In stream-based sampling one selects one item at a time and decides whether it is worth the cost of classifying it \cite{atlas1990stream}. In pool-based sampling examples are sampled from a large pool of unlabelled data \cite{wang2015active}. As we will see, our algorithm can be used both as a pool-based sampling or as query synthesis, depending on the points used.
The most typical strategies to select the samples are \textit{uncertainty sampling}, that selects points with maximum uncertainty about which class they belong to \cite{lewis1994uncertainty}, or \textit{Query-by-committee}, where the space of classifiers that agree with the data is halved sequentially \cite{melville2004commitee}. Finally there is also the strategy of using \textit{expected error reduction} \cite{roy2001error} that selects those points that on average, weighted according to probability, make the loss function as low as possible. This last procedure is similar to what we are using, except that instead of minimizing the loss function, we select points that on expectation would achieve the margin of the SVM to be as low as possible.

Finally, let us make a brief introduction to quantum algorithms used for quantum machine learning, since the quantum SVM that we have used through the text is just one of the earlier works on this field. The possibility of applying quantum techniques in the machine learning and artificial intelligence domain have attracted in the last decade much interest, and a review of this efforts is \cite{dunjko2018machine}. For example, following work on quantum SVM, various kernel methods have been developed to improve Support Vector Machines, such as \cite{schuld2019quantum} or \cite{perez2020data}. Most of this efforts have been focused on approaches that can be implemented NISQ devices. For instance, the later reference, \cite{perez2020data}, was specially focused on constructing a universal quantum classifier with minimal amount of quantum resources, using a technique called data re-uploading.

Other machine learning techniques that have been quantized include Bayesian learning \cite{zhao2019bayesian}, where a polynomial speedup is expected thanks to the use of linear algebra techniques such as variations over HHL.

Finally, although considerable effort has been put on quantizing machine learning algorithms such as Support Vector Machines and kernel methods, given the success of the Deep Learning (aka neural networks) techniques in tackling many problems that include image recognition and text translation, significant effort has be put in quantizing them. One example of this is \cite{farhi2018classification}, or the recent publication of TensorFlow Quantum \cite{tensorflow_quantum}, the TensorFlow library to work with quantum neural networks. One particular example of a Deep Learning technique that has received much attention both in the classical and quantum machine learning communities are the generative adversarial networks \cite{zoufal2019quantum}. This networks are composed of two parts, a generator and a discriminator that compete, and a distribution $\mathcal{D}$. The generator tries to generate a distribution resembling as much as possible $\mathcal{D}$, and the discriminator tries to differentiate between the actual $\mathcal{D}$, and the one generated by the generator. The game ends when the generator has fully learned $\mathcal{D}$, and precisely this is the reason why adversarial generative networks have been devised as a method to prepare quantum states efficiently.

However, we would like to emphasise that although somewhat related to adversarial examples via adversarial training, the setup in the later case is different. For example, in the setting of adversarial examples our classifier does not need to know how to generate any state of the distribution, but only how to classify an input in one of the classes. In such respect, there is no generative device in adversarial examples, except perhaps the adversary who is trying to fool the classifier, whereas in the generative adversarial training setup there is a single distribution, and therefore a single class. Perhaps more importantly, adversarial examples seem to be a general and somewhat unfortunate feature of many different kinds of classifiers, whereas generative adversarial networks are a particular, although useful, technique.

\section{\label{sec:active_learning}Modelling the active learning problem}

As mentioned in the introduction, the problem we want to solve in this article is the following: 
\begin{Problem}
\textbf{Informative Points Sampling Problem} \label{Informative sampling problem}
Suppose we are given a set of points $S=(\vec{x}_i, \vec{c}_i)$, $i \in \{1,..,n\}$, where $\vec{x}_i \in \mathbb{R}^m$ are the $n$ data points already classified, and $\vec{c}_i$ are label vectors that indicate probability of membership to the possible classes where the point might be classified. Suppose also that these memberships sum up to, at most, 1 for each point, and for simplicity assume that there are just two classes. The problem is: what are the most informative points to be added to $S$ in order to learn the Support Vector Machine, and in particular the minimum distance between classes $r_p$, with better precision?
\end{Problem}

This problem makes reference to two key concepts: Support Vector Machine and `informativeness', that we introduce now. Although the technical definition of Support Vector Machine (SVM) is stated in the section \ref{sec:background_SVM}, it is basically the hyperplane of separation between two classes that maximizes the margin.

In our particular problem, we are interested in finding the minimum distance between two classes, in order to provide the required covering of the two classes that avoid having adversarial attacks. However, it might be the case that both classes have some points in common, e.g. $(\vec{c}_i)_j, (\vec{c}_i)_k> 0$, where $i$ indicates the point and $j$ and $k$ two classes. In order to avoid this, we establish that a point $\vec{x}_i$ is in class $j$ if $(\vec{c}_i)_{j} > 0.8$, and as $\sum_j (\vec{c}_i)_j = 1 $ for any point $\vec{x}_i$, this will avoid any overlap between classes. This value is arbitrary but should be strictly greater than $0.5$. This way we clearly separate the two sets, and the minimum distance between classes is greater than $0$. With this condition one wants to make classes clearly separated, since otherwise it does not make sense the concept of adversarial example, which will be central to our discussion. If there are more than two classes, then one can generalise the previous setting making comparison on all possible pairs of classes, which scales quadratically with the number of classes.

\begin{figure}
    \centering
    \includegraphics[scale=.75]{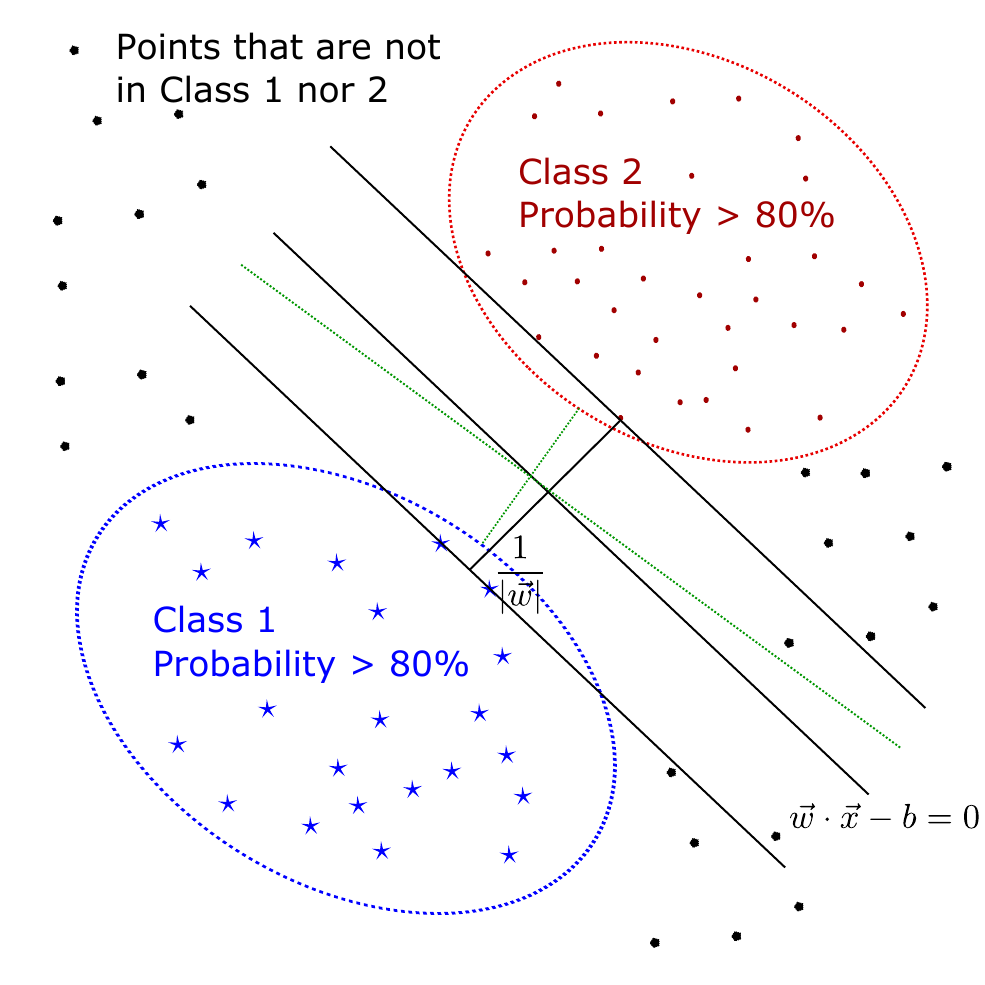}
    \caption{Example of a SVM, characterised by the equation $\vec{w}\cdot \vec{x}-b=0$. The points depicted are the initial training set, with colors according to their classification, and those in black are those not belonging to any class according to our criteria that for a point $\vec{x}_i$ to be in class $j$, $c_{ij}>0.8$. Initially we have the SVM in black, but we want to obtain the green one, that is more accurate. The dotted lines are the ones that we cannot see initially and must find out. The equation of the SVM is the one depicted and the margin $1/|\vec{w}|$ is chosen to be equal to $r_p$. This is achieved when we make the two parallel hyperplanes that indicate the margin to fulfill equations $\vec{w}\cdot \vec{x}-b=\pm 1$. Since making the SVM more precise implies making the margin smaller, we want to find an SVM that maximizes $|\vec{w}|$.}
    \label{fig:SVM}
\end{figure}

Now let us look at figure \ref{fig:SVM}. We can see that the initially calculated SVM has a greater margin than that of the SVM we would have if we were able to perfectly know both classes. If we used such margin in theorem \ref{Teorema 5}, we would not find a cover that avoids adversarial examples. Thus, we are interested in minimizing the maximum margin, which is given by the true SVM if we knew the actual classes and not just some samples. 

Our problem is however in contrast with usual pool-based sampling (see section \ref{sec:related} for more information). In their problem the algorithm usually seeks to classify new points near the SVM. This is not the case for our problem, since we want points that have a membership to a class greater than $0.8$. Points that have no clear membership to a single class are not so interesting and we do not include them in our data set. 

Now we turn to the concept of `informativeness'. How then do we measure how interesting could be to classify an arbitrary point? A good heuristics for our problem is that we are trying to find points that with high probability would decrease the margin $1/|\vec{w}|$ a lot, to get a better estimate of $r_p$. This is because there are two competing conditions on $r_p$. On the one hand, in order to fulfill the condition of the problem, we cannot take $r_p$ larger than it really is, as that would make us choose a cover of the classes that does not fulfill the conditions of the theorem 5 of \cite{Geometry_of_adversarial_examples}. On the other, the smaller $r_p$ is, the more expensive it is to establish the cover of the classes.

The previous paragraph suggests measuring the `informativeness' of a point by dividing the problem in two parts: we first find the probability that a given point $\vec{x}_{n+1}$ is in class $c$, and then multiply this probability by the inverse of the margin distance of the updated SVM, taking into account this would-be newly classified point, $|\vec{w}_{n+1}|$.

The `informativeness' will thus measure an expected value of information gain if a given point was in a given class. As such, it will be the product of a probability of $\vec {x}$ being in class $c$, $P_c(\vec{x})$, times the information we gain, that we will measure as $|\vec{w}|$. The reason for this choice is that we want to add points to training set, that give a more precise account of the classes. That is, we wish to minimize the margin of size $1/|\vec{w}|$, and therefore we want to maximize $|\vec{w}|$.

\begin{Definition}
\textit{`Informativeness'} measures the expected value of information we get by adding a point $\vec{x}_i$ to the training set. We will therefore define it as the product
\begin{equation}
    P_c(\vec{x}_i) \cdot |\vec{w}_{\vec{x}_i}|,
\end{equation}
where $P_c(\vec{x}_i)$ is the probability that $\vec{x}_i$ is in class $c$, and $1/|\vec{w}_{\vec{x}_i}|$ is the size of the margin of the resulting SVM if $\vec{x}_i$ really were in such class.
\end{Definition}

\section{\label{sec:background_SVM}Background on Support Vector Machines}

Since we will rely on them somewhat heavily, in order to carry out those results it is useful to remember some results from \cite{QSVM}, that explains how a quantum Support Vector Machine algorithm works. Let us first introduce the definition of a Support Vector Machine.

\begin{Definition} \textbf{Support Vector Machine (SVM)}
Let $\mathcal{M}$ be a $m$-dimensional manifold with two submanifolds or classes and a set of points already classified $\{\vec{x}_i,y_i\}$, $\vec{x}_i$ being the point and $y_i$ the label. Then a \textit{Support Vector Machine} is a hyperplane in the manifold separating both classes such that minimum distance between $\{\vec{x}_i\}$ and the hyperplane is as large as possible. If the SVM is linear it may be described by equation
\begin{equation}
    \vec{w}\cdot \vec{x} - b = 0,
\end{equation}
and the size of the margin is $1/|\vec{w}|$, which would be equal to $r_p/2$ if the SVM was perfect. One may also define a non-linear SVM using a non-linear kernel for the dot product.

We highlight that the name Support Vector Machine (SVM) will refer to both the algorithm and the separation hyperplane, the decision boundary.
\end{Definition}

The authors of \cite{QSVM} assume that each point $\vec{x}_i$ is labeled with a single class $y_i$, and there are only two classes $y_i = \pm 1$. Given the pairs of data and label $(\vec{x}_i,y_i)_{i\in \{1,...,n\}}$, the authors state that calculating the SVM is equivalent, in the dual formulation, to maximizing over the multipliers $\vec{\alpha}$ of the Lagrangian
\begin{equation}
    L(\vec{\alpha})=\sum_{i=1}^n \alpha_i y_i -\frac{1}{2}\sum_{i,k=1}^n \alpha_i K_{ik} \alpha_k,\label{lagrangian}
\end{equation}
with constraints $\sum_i \alpha_i = 0$ and $\alpha_i y_i \geq 0$ $\forall i$. After this, the result of the classifier is given, for a new point $\vec{x}$ as
\begin{equation}
    y(\vec{x})=\text{sgn}\left(\sum_{i=1}^n\alpha_i \sum_{l,k=1}^m K_{l,k}x_{i,l} x_k+b\right),
\end{equation}
with $K_{l,k}=\vec{x}_l \cdot \vec{x}_k$, and may be rewritten as 
\begin{equation}
    y(\vec{x})= \text{sgn} \left(K(\vec{w},\vec{x})+b\right),\quad \vec{w}:= \sum_i \alpha_i \vec{x}_i. \label{w definition}
\end{equation}
Notice that the matrix $K_{l,k}$ is a kernel matrix that defines a dot product. Since the margins are of length at least 1, this means that for the training data
\begin{equation}
    y_i(\vec{w}\cdot \vec{x}_i+b)\geq 1. \label{hard margin equation}
\end{equation}
 Since there are only two classes, $y_i=\pm 1$, and so $y_i^2=1$. We will later say, with a bit of abuse of notation, that a given point is in class $c$ to mean any of the two possible values of $y_i$. Additionally, we can transform the previous inequality results into equality adding slack variables $e_i$, 
\begin{equation}
    (\vec{w}\cdot \vec{x}_i+b)=y_i-y_i e_i,
\end{equation}
such that if we allow $e_i>0$, we will be allowing for a soft margin, that is, some points may not fulfill \eqref{hard margin equation}. In such case we will add to Lagrangian \eqref{lagrangian} a term $(\gamma/2) \sum_j e_j^2$, where $\gamma$ is specified by the user, to penalise any violation of \eqref{hard margin equation}.
Then, minimising the Lagrangian is equivalent to solving the following least square approximation \cite{QSVM},
\begin{equation}
    F
    \begin{pmatrix}
    b\\
    \vec{\alpha}
    \end{pmatrix}
=
\begin{pmatrix}
0 & \vec{1}^T\\
\vec{1} & K+\gamma^{-1}\mathbf{1} 
\end{pmatrix}
    \begin{pmatrix}
    b\\
    \vec{\alpha}
    \end{pmatrix}
    =    \begin{pmatrix}
    0\\
    \vec{y}
    \end{pmatrix}.
    \label{system of equations}
\end{equation}
Here is where \cite{QSVM} uses quantum linear algebra techniques to solve \eqref{system of equations}, as we will see later, and obtain the result $\ket{b,\vec{\alpha}}$.
To classify a new point $\ket{\vec{x}}$, one constructs, making use of qRAMs,
\begin{equation}
    \ket{\Tilde{u}}:=\frac{1}{\sqrt{N_{\Tilde{u}}}}
    \left(b\ket{0}\ket{0}+\sum_{k=1}^n\alpha_k|\vec{x}_k|\ket{k}\ket{\vec{x}_k}\right)
    \label{u tilde}
\end{equation}
and 
\begin{equation}
    \ket{\Tilde{x}}:=\frac{1}{\sqrt{N_{\Tilde{x}}}}
    \left(\ket{0}\ket{0}+\sum_{k=1}^n|\vec{x}|\ket{k}\ket{\vec{x}}\right).
    \label{x tilde}
\end{equation}
Remember that qRAMs perform $\sum_k\beta_k\ket{k}\rightarrow \sum_k\beta_k\ket{k}\ket{\vec{x}_k}$, for $\beta_k$ arbitrary amplitudes. Using the qRAM proposed in \cite{QRAM_bucket_brigade} only $O(\log n)$ gates are activated, although $O(n)$ should be present in the circuit.

If the previous is possible, one should also be able to prepare the state $\ket{\psi}= 1/\sqrt{2}(\ket{0}\ket{\Tilde{u}}+\ket{1}\ket{\Tilde{x}})$ and measures the probability of the ancilla being in state $\ket{-}=1/\sqrt{2}(\ket{0}-\ket{1})$,
\begin{equation}
    P=\frac{1}{2}(1-\braket{\Tilde{u}|\Tilde{x}}). \label{probability of 0}
\end{equation}
This is equivalent to performing a Hadamard gate over the first ancilla register and measuring the probability of obtaining $\ket{1}$, and is called Swap Test \cite{SupervisedQML}. If $P\geq 1/2$, $y(\vec{x})=1$, otherwise $y(\vec{x})=-1$.

\section{\label{sec:Main algorithm} Main algorithm}

As we have seen, our aim is to sample points to be added to the training set, $\vec{x}_{n+1}$, with the highest possible informativeness, defined as $P_c(\vec{x}_{n+1})|\vec{w}_{n+1}|$, where the first term indicates the probability that a given point is in a class, and the second measures how much that would improve the classifier.

Thus, we will employ the strategy explained in the Algorithm \ref{Main algorithm} to sample from the $1/C$ most relevant points.

\begin{figure}
\begin{algorithm}[H]
\caption{Active learning against adversarial examples.}\label{Main algorithm}
\begin{algorithmic}[1]
\Procedure{Active learning against adversarial examples}{}
\BState \textbf{To find a point $\vec{x}$ that is in the $1-1/C$ quantile of `informativeness' with probability $1-e^{-\beta}$, iterate $O(C\beta)$ times:}
\State Uniformly at random sample a point $\vec{x}_{n+1}$ in the space.
\State Evaluate, as explained in section \ref{sec:Probability}, the probability that a point is in a given class $c$. We get $P_c(\vec{x}_{n+1})$. \label{step probability}
\State Solve the linear system of equations \eqref{system of equations} using the procedure of \cite{QSVM}, to obtain the state $\ket{b,\vec{\alpha}}_{n+1}$. It is explained in section \ref{sec:QSVM}. \label{step QSVM}
\State Perform \eqref{second matrix system}, using the Chebyshev approach taken from \cite{QLSAchilds}, to calculate $\ket{\vec{w}_{n+1}}$. Then use Amplitude Estimation to estimate $|\vec{w}_{n+1}|$. The procedure to calculate $\ket{\vec{w}_{n+1}}$ is described in section \ref{sec:norm}, and its norm in appendix \ref{sec:Norm_appendix}. \label{step norm}
\State Calculate the informativeness $P_c(\vec{x}_{n+1})\cdot |\vec{w}_{n+1}|$ of point $\vec{x}_{n+1}$. If it is higher than the previous best `informativeness', save the pair $(\vec{x}_{n+1},P_c(\vec{x}_{n+1})\cdot |\vec{w}_{n+1}|)$ to memory substituting the previous best pair of values.
\State After the $O(\beta C)$ iterations, output the saved best pair.
\EndProcedure
\end{algorithmic}
\end{algorithm}
\end{figure}

\subsection{\label{sec:Probability}Calculating $P_c(\vec{x}_{n+1})$}

The first thing we should care about is calculating $P_c(\vec{x}_{n+1})$, the probability that point $\vec{x}_{n+1}$ is in class $c$. 

The simplest way to calculate the probability would be to solve the SVM, calculate the distance from the point $\vec{x}_{n+1}$ to the decision boundary, and apply an activation function that converts the distance to a probability.
Solving the SVM can be done using \cite{QSVM} and reading each entry using Amplitude Estimation. It would return the vector $(b,\vec{\alpha})$, which can be used to create $\vec{w}=\sum_j \alpha_j \vec{x}_j$, and therefore the SVM.

A second, more elegant solution, is the following. In \cite{QSVM}, the authors propose a method to estimate, using the Swap Test and the output state $\ket{b, \vec{\alpha}}$ , to which class does a given point $\ket{\vec{x}}$ belong. They calculate that the success probability of measuring a $\ket{-}$ in the ancilla is
\begin{equation}
    P=\frac{1}{2}(1-\braket{\Tilde{u}|\Tilde{x}}),
\end{equation}
for $\ket{\Tilde{u}}$ and $\ket{\Tilde{x}}$ defined as in \eqref{u tilde} and \eqref{x tilde}.
The interesting thing to notice is that if $P>1/2$ the classification is in one class and if $P<1/2$ it is in the other. We modify this protocol slightly so that we perform Amplitude Estimation on this result instead of repeatedly measuring the expected value, which improves the complexity from $O(\epsilon^{-2})$ to $O(\epsilon^{-1})$. This allows us to obtain the amplitude $A_{\vec{x}_{n+1}} = \sqrt{P}$. Then, one may use $P$ as $P_c(\vec{x}_{n+1})$, or apply whatever activation function we consider appropriate to shape $P_c(\vec{x}_{n+1})$, like a sigmoid for example.

\subsection{\label{sec:QSVM} Quantum Support Vector Machine}

Now we focus on the main subroutine that our algorithm uses, which is mostly based in \cite{QSVM}, and that outputs $\ket{b,\vec{\alpha}}$, the solution to \eqref{system of equations}. This subroutine will be key to calculating $|\vec{w}_{n+1}|$, as one can use \eqref{w definition} to calculate $\vec{w}$, whose norm we will estimate later. In this section we will review how \cite{QSVM} performs Hamiltonian Simulation, and its role in the HHL algorithm \cite{HHL}, as well as some variations of that algorithm. We will also show the low rank approximation needed to achieve the exponential speedup, and the role of the condition number as a normalization factor in the estimation of $||(b,\vec{\alpha})||$. Finally, we will review the complexity of this calculation, which relies on the use of qRAMs to achieve an exponential speedup.

Thus, the first step is solving \eqref{system of equations}, which \cite{QSVM} does by using HHL algorithm \cite{HHL} with a different Hamiltonian simulation. This is due to the fact that in general the kernel matrix $K$ is dense, and the Hamiltonian simulation of \cite{HHL} is better suited for the sparse case.

Rather, the authors of \cite{QSVM} use a technique they had previously developed in \cite{QPCA} that allows for efficient Hamiltonian simulation of dense low-rank matrices. Let us explain how to do it. Suppose we have a given quantum state $\sigma$, and we are able to prepare $T$ copies of a density matrix $\rho$. One would like to perform the Hamiltonian simulation
\begin{equation}
    e^{-i t \rho}\sigma e^{i t \rho}. \label{matrix application}
\end{equation}
Repeated application of 
\begin{equation}
    \mathrm{tr}_1(e^{-i \Delta t S}\rho\otimes\sigma e^{i \Delta t S}) = \sigma -i \Delta t [\rho, \sigma] + O(\Delta t^2), \label{low rank hamiltonian simulation}
\end{equation}
$S$ the Swap operator, approximates
\begin{equation}
    e^{-i T\Delta t \rho}\sigma e^{i T\Delta t \rho},
\end{equation}
$T$ indicating the number of times we apply \eqref{low rank hamiltonian simulation}.
The authors of \cite{QPCA} observe that to simulate \eqref{matrix application} with precision $\epsilon^{-1}$ one has to repeat the process $ O(t^2\epsilon^{-1})$ times, each taking time $O(\log m)$ since that is the time it takes to prepare $\rho$ using a qRAM.

In our case $\rho$ is the matrix $F$, that will be represented by operator $\hat{F} = (J + K + \gamma^{-1} \mathbf{1})/ \mathrm{tr} F$, where
\begin{equation}
    J = \begin{pmatrix}
    0 & \vec{1}^T\\
    \vec{1} & 0
    \end{pmatrix}.
\end{equation}
Then, $e^{i\Delta t \hat{F}}= e^{i\Delta t J/\mathrm{tr} F}e^{i\Delta t K/\mathrm{tr} F}e^{i\Delta t \gamma^{-1} \mathbf{1}/\mathrm{tr} F}$. Simulating $J$ is described in \cite{childs2010relationship}, and the matrix $\gamma^{-1} \mathbf{1}$ is also easy. An important insight of \cite{QSVM} is how to prepare a state with density matrix $K/\mathrm{tr} F$. If we were able to prepare $K/\mathrm{tr} K$ then one would only need to correct for a scalar term $\mathrm{tr} K/ \mathrm{tr} F = O(1)$ in the simulation time. In the appendix A of \cite{QSVM} it is explained how to estimate $\mathrm{tr} K$, which can be used to estimate $\mathrm{tr} F = \mathrm{tr} K + \gamma \mathrm{tr}  \mathbf{1}$.

So, let us show how \cite{QSVM} prepares a density matrix state $K/\mathrm{tr} K$. To do so, prepare, using a qRAM:
\begin{equation}
    \ket{\chi} = \frac{1}{\sqrt{N_\chi}}\sum_i |\vec{x}_i| \ket{i}\ket{\vec{x}_i},
\end{equation}
Tracing out the second register prepares
\begin{equation}
\begin{split}
    \mathrm{tr}_2 (\ket{\chi}\bra{\chi}) &= \frac{1}{N_\chi} \sum_{i,j=1}^n\braket{\vec{x}_i|\vec{x}_j}|\vec{x}_i||\vec{x}_j|\ket{i}\bra{j}\\
    &= K/\mathrm{tr} K.
\end{split}
\end{equation}

The previous method allows to prepare the density matrix $K/\mathrm{tr} K$ needed to implement the Hamiltonian simulation of \eqref{low rank hamiltonian simulation}, in time complexity $O(\epsilon^{-1}t^2 k)$. Therefore, we can perform the Hamiltonian simulation necessary to implement the HHL algorithm efficiently.

The HHL algorithm consists on the following steps. First, one formally decomposes the state $\ket{0,\vec{y}}$ of \eqref{system of equations} in eigenvectors $\ket{0,\vec{y}} = \sum \beta_j \ket{u_j}$ of the Hamiltonian defined by matrix $F$ in \eqref{system of equations}. Next, one Phase Estimates the eigenvalues, using a Hamiltonian simulation algorithm such as the one explained above, obtaining $\sum \beta_j \ket{u_j}\ket{\lambda_j}$. Finally, one performs the controlled rotations
\begin{equation}
\begin{split}
    &\sum_j \beta_j \ket{u_j}\ket{\lambda_j}\ket{0}\rightarrow \\
    &\sum_j \beta_j \ket{u_j}\ket{\lambda_j}\left(\frac{1}{\lambda_j\kappa}\ket{1}+\sqrt{1-\frac{1}{\lambda_j^2\kappa^2}}\ket{0}\right),
\end{split} \label{HHL}
\end{equation}
and postselects in ancilla state $\ket{1}$ to obtain the solution to the system of equations \cite{HHL}.

The stated complexity of the original HHL algorithm is $O(\kappa^2 \epsilon^{-1}\text{poly}\log n)$, $\kappa$ the condition number. The complexity of the originally used Hamiltonian simulation algorithm is $O(d\epsilon^{-1}\text{poly}\log n)$, where $d$ the sparsity, for the sparse oracle access model \cite{HHL}. The complexity of the condition number has been lowered to $O(\kappa)$, as shown using the technique of Variable Time Amplitude Estimation in \cite{Ambainis}, which is optimal in this parameter, also proved in that article. On the other hand, relying on more efficient Hamiltonian simulation techniques, \cite{QLSAchilds} proposed a variation of the HHL algorithm that does not require Amplitude Estimation thus reducing the complexity from $O(\epsilon^{-1})$ to $O(\text{poly}\log \epsilon^{-1})$. Unfortunately, the Hamiltonian simulation described above still has complexity $O(\epsilon^{-1})$, making it impossible to reduce the complexity of the overall method further.

There is an alternative to the Hamiltonian simulation method that we have described. In \cite{Dense_hamiltonian_simulation}, a technique is introduced that allows to simulate a dense Hamiltonian with complexity $O(\sqrt{n}\text{poly}\log \epsilon^{-1})$, for Hamiltonians of any rank, and it can be used with the techniques from \cite{QLSAchilds}, to reduce the complexity on the precision from linear to polylogarithmic. The caveat is that it requires to use a special quantum-accessible data structure that is explained in the appendix \ref{sec:definitions} and plays the role of a Quantum Read-Only Memory, and has the complexity stated above, polynomial in $n$.

Coming back to our main discussion, once we have the state $\ket{b,\vec{\alpha}}$, we would like to recover the norm of vector $\vec{w}$. As a first step, we would like to obtain the norm $||(b,\vec{\alpha})||$. Suppose we are trying to solve $Ax=b$, where the largest eigenvalue of $A$ is less or equal to 1; else see the end of appendix \ref{sec:Norm_appendix} for a minor correction. As explained in appendix A of \cite{FEM}, we can calculate the norm of the solution $||x||$ using
\begin{equation}
    ||x||= \kappa ||b|| \sqrt{p_1}, \label{norm solution HHL}
\end{equation} 
where $\sqrt{p_1}$ represents the amplitude of the postselection ancilla of HHL algorithm being in the correct state, usually $\ket{1}$. This ancilla is used to perform the non-unitary part of the algorithm via a measurement. The reason for the previous equation \eqref{norm solution HHL} is because the acceptance probability of HHL scales as
\begin{equation}
    p_1 = \frac{||A^{-1}b||}{||b||\kappa^2}.
\end{equation}
Estimating such amplitude $\sqrt{p_1}$ thus requires of Amplitude Estimation \cite{Amplitude_estimation}, with cost $O(\epsilon^{-1})$. 

Now let us turn to the condition number of the system of equations that appears in the SVM we are solving, the condition number of the matrix $F$ in \eqref{system of equations}. Recall that the condition number is defined as
\begin{equation}
    \kappa= \frac{\sigma_{\max}}{\sigma_{\min}},
\end{equation}
where $\sigma_{\min}$ and $\sigma_{\max}$ are the minimum and maximum singular values respectively. The condition number will be important to correct the norm of the solution to the system of equations, as can be seen from \eqref{norm solution HHL}.
However, calculating it with the Quantum Singular Value estimation technique from \cite{kerenidis2017recommendation} might be too expensive. 

On the other hand, the authors of the quantum SVM article, \cite{QSVM}, propose that in the case where the kernel matrix has $O(1)$ eigenvalues of size $O(1)$, and $O(n)$ eigenvalues with values $O(1/n)$ as it is in our case, we can choose a condition number $\kappa_{\mathrm{eff}}=O(1)$ such that in the end we will get an additional error of order $O(1/\sqrt{n})$, in addition to $\epsilon$.  

This low rank approximation means that the algorithm in \cite{QSVM} only takes into account the eigenvalues $\lambda_i$ that are $\epsilon_K\leq \lambda_i \leq 1$. The main idea of the low rank approximation is filtering out those eigenvalues $\lambda< \kappa_{\mathrm{eff}}^{-1}$, so the final rotation of the HHL algorithm, indicated in \eqref{HHL}, is performed only if $\lambda_j > \kappa_{\mathrm{eff}}^{-1}$ and imposes $\kappa = \kappa_{\mathrm{eff}} = O(1)$.

In appendix C of the supplementary material of \cite{QSVM} the authors show that $||K-K_{\mathrm{eff}}|| = \sqrt{\sum_{\lambda_i = O(1/n)}\lambda_i^2}$, with $K_{\mathrm{eff}}$ the `filtered' low-rank approximation of $K$, $K_{\mathrm{eff}} = \sum_{\lambda_i \geq \kappa_{\mathrm{eff}}^{-1}}\lambda_i \ket{u_i}\bra{u_i}$. Since there are $O(n)$ eigenvalues of size $O(1/n)$, the induced error is of order $O(n^{-1/2})$. Then, one can use theorem 1 in the appendix of the HHL algorithm, \cite{HHL}, to show that, if $\ket{b,\vec{\alpha}}$ is the exact solution of the system of equation; and $\ket{b,\vec{\alpha}}_{\mathrm{eff}}$ the approximate one after postselecting on the subspace spanned by the eigenvalues $\lambda_j \geq \kappa_{\mathrm{eff}}^{-1}$ and the ancilla in state $\ket{1}$, then $||\ket{b,\vec{\alpha}}_{\mathrm{eff}}-\ket{b,\vec{\alpha}}|| = O(\epsilon +n^{-1/2})$. Thus, for relatively large $n$ the induced error with this approximation is small.

Overall, this implies that instead of \eqref{norm solution HHL}, we will have 
\begin{equation}
    ||x||= \kappa_{\mathrm{eff}} ||b|| \sqrt{p_1}, \label{norm solution QSVM}
\end{equation} 
In such case notice that we have imposed an effective condition number for all candidate points $\vec{x}_{n+1}$: $\kappa_{\vec{x}_{n+1}}=\kappa_{\mathrm{eff}}$, so we no longer have to care about the condition number: it will be the same scale factor for all points $\vec{x}_{n+1}$, and thus without relevance. The same will happen also for $||b||$ in \eqref{norm solution QSVM} since $b = (0, y_1,...,y_{n+1})^T$, with $y_{n+1}$ the same for all candidate points.
Since we suppose $n$ is large enough, the additional additive error we introduce will be small, of order $O(n^{-1/2})$ according to the appendix C in the supplementary material of \cite{QSVM}. Thus, the complexity of the algorithm is $O(\epsilon^{-3}\kappa_{\mathrm{eff}}^3\log(mn))$, where $m$ is the dimension of the space. 

To finish this section, let us recall how to calculate the complexity of the algorithm of \cite{QSVM}. They claim that \eqref{low rank hamiltonian simulation} has error $\epsilon = \Tilde{O}(||\hat{F}||^2\Delta t^2 )$, $||\hat{F}||$ the Frobenius norm of $\hat{F}$, and it is repeated $T$ periods. Thus, $\epsilon =\Tilde{O}(||\hat{F}||^2 \Delta t^2 T)= \Tilde{O}(||\hat{F}||^2 t^2/T)$, for $\Delta t = t/T$. This implies a simulation cost $T = \Tilde{O}(t^2 ||\hat{F}||^2 \epsilon^{-1})$, when implementing \eqref{low rank hamiltonian simulation} is taken at unit cost.

On the other hand, HHL requires phase estimating the eigenvalues, so the relative error of $\lambda^{-1}$ can be indicated as $\epsilon = O(1/\lambda t)\leq O(\kappa_{\mathrm{eff}}/t)$. Therefore $t = O(\kappa_{\mathrm{eff}} \epsilon^{-1})$, and substituting in the previous paragraph, $T = O(||\hat{F}||^2 \kappa_{\mathrm{eff}}^2 \epsilon^{-3} \log(mn))$. An additional $\kappa_{\mathrm{eff}}$ is needed to perform postselection on the result. Thus, the overall complexity of the algorithm is $O(o ||\hat{F}||^2  \kappa_{\mathrm{eff}}^3\epsilon^{-3}\log(mn))$, for a kernel of order $o$.

If instead of using the basic HHL technique we had decided to use more advanced ones, such as the Fourier approach described in \cite{QLSAchilds} (see appendix \ref{sec:definitions}), the complexity would have been $\Tilde{O}(T\alpha)$, for $T = \Tilde{O}(||\hat{F}||^2t^2\epsilon^{-1})$, $t = \Tilde{O}(\kappa)$ and $\alpha = \Tilde{O}(\kappa)$, ignoring polylogarithmic factors \cite{QLSAchilds}. Overall, the complexity would be $\Tilde{O}(||\hat{F}||^2\kappa^3\epsilon^{-1})$.

\subsection{\label{sec:norm}  The norm of $|\vec{w}_{n+1}|$}

The aim of the previous section was to obtain $\ket{b,\vec{\alpha}}$ and its norm. The aim of this one is to calculate $\vec{w}$ from that result. We will see that one may do that using a Fourier expansion, at cost $O(\epsilon^{-1})$; or better, use of Chebyshev series as in \cite{QLSAchilds}, at cost polylogarithmic in all variables, provided the use of qRAMs. In appendix \ref{sec:Norm_appendix} we explain how to calculate the norm of $|\vec{w}_{n+1}|$ from $\vec{w}$, which only requires performing Amplitude Estimation once, and normalizing with some factors.

The first, trivial, idea we had to calculate $\ket{\vec{w}_{n+1}}$ is reading all the entries of $\ket{b,\vec{\alpha}}$ using Amplitude Estimation. Then one may use classical computing to compute $\vec{w}= \sum_{i=1}^{n+1} \alpha_i \vec{x}_i $, and finally its norm $|\vec{w}_{n+1}|= \sqrt{\sum_{i=1}^m (\sum_{j=1}^{n+1} x_{ij,x_{n+1}} \alpha_{j,x_{n+1}})^2}$. Amplitude Estimation does not immediately recover the sign of each $\alpha_i$, but finding it is not complicated either. Once we know $|\alpha_i|$ and $|\alpha_j|$ we can prepare the state $C_{ij} (|\alpha_i| \ket{j} + |\alpha_j| \ket{i})$ and perform a Swap Test with the solution vector $\ket{b, \vec{\alpha}}$. If the relative sign of entries $i$ and $j$ is the same we will get a nonzero result proportional to $2C_{ij} \alpha_i \alpha_j$, but if the relative sign is opposite, the dot product will cancel out \cite{QIP}. Establishing the relative sign of two entries is enough, since the norm will not care about the global sign.

However, the procedure described above is time consuming, as one needs to prepare the solution of the system of equations \eqref{system of equations} $O(n+1)$ times in order to read all the entries of the solution and calculate their relative sign. Also, calculating $\vec{w}$ takes complexity $O(nm)$ using classical computing, so, instead of that we will try to prepare $\ket{\vec{w}_{n+1}}$ from $\ket{b,\vec{\alpha}}$, and estimate the change in the norm of the result. Notice in the first place that the Quantum SVM article \cite{QSVM} is able to classify a point without first calculating $\ket{\vec{w}_{n+1}}$, as indicated in \eqref{probability of 0} and subsequent paragraph. Also, it is not clear how to prepare $\ket{\vec{w}_{n+1}}$ from the state in \eqref{u tilde}. 

Instead, since $\vec{w}= \sum_i \alpha_i \vec{x}_i$, this suggest multiplying the solution vector from the Quantum SVM, $\ket{b, \vec{\alpha}}$ by a matrix operator whose entries are $x_{i,j}$. Explicitly, 
\begin{equation}
\begin{split}
A\ket{b,\vec{\alpha}} = 
\begin{pmatrix}
0 & x_{1,1} & ... & x_{1,m}\\
\vdots &    &     & \vdots \\
0 & x_{n+1,1} & ... & x_{n+1,m}
\end{pmatrix}
\begin{pmatrix}
b\\
\alpha_1\\
\vdots \\
\alpha_{n+1}
\end{pmatrix}\\
= 
\begin{pmatrix}
\sum_{i=1}^m  \alpha_j x_{1,j}\\
\vdots \\
\sum_{i=1}^m  \alpha_j x_{n+1,j}\\
\end{pmatrix}
= \vec{w}.
\end{split} \label{second matrix system}
\end{equation}
Performing this operation classically, requires $O(nm)$ operations. Can we perform this operation in a quantum way?

Our strategy to calculate the previous matrix-vector product will take inspiration from the algorithms presented in \cite{QLSAchilds} to simulate $A^{-1}$. 
The fact that in our case $A$ is not necessarily square should not pose any problem, since one can write
\begin{equation}
\begin{blockarray}{ccc}
 & m+1 &  n+1 \\
\begin{block}{c(cc)}
     m+1& \mathbf{0} & A^T   \\
     n+1 & A & \mathbf{0} \\
\end{block}
\end{blockarray}
    \quad     
    \begin{blockarray}{c}
    \\
    \begin{block}{(c)}
    b\\
    \vec{\alpha}\\
    \vec{0}_{n+1}\\
    \end{block}
    \end{blockarray}\quad = \quad
    \begin{blockarray}{c}
    \\
    \begin{block}{(c)}
    \vec{0}_{m+1}\\
    \vec{w}\\
    \end{block}
    \end{blockarray}\quad.
    \label{new Matrix system}
\end{equation}
Let us call $M$ the matrix from this previous equation. The first thing we have to do is to attach a register to $\ket{b,\vec{\alpha}}$, that we set to $\ket{0}$ for the entry $b$ (that is, the original register is in state $\ket{0}$) and $\ket{1}$ otherwise. This will allow us to substitute any of the $0$s in the matrix in \eqref{second matrix system}, but avoiding interference of the entry of $b$ with those of $\vec{\alpha}$. The aim of this is to avoid $M$ being ill-conditioned, although it is only a technical detail. Then all the singular values fall in $[1/\kappa_M, 1]$, and we can rewrite
\begin{equation}
    \begin{pmatrix}
    b\\
    \vec{\alpha}\\
    \vec{0}_{n+1}
    \end{pmatrix}
    \rightarrow
    \begin{pmatrix}
    b\\
    \vec{0}_{m}\\
    \vec{0}_{n+1}
    \end{pmatrix}
    \otimes \ket{0}
    +
        \begin{pmatrix}
    0\\
    \vec{\alpha}\\
    \vec{0}_{n+1}
    \end{pmatrix}\otimes\ket{1}.
\end{equation}
Notice that by linearity of quantum mechanics the unitary we were going to apply to the the left-hand side is applied individually to each of the two terms in the decomposition. Later we will only care about the case when the additional ancilla we have added is in state $\ket{1}$.

With this remark, the question becomes twofold: how to decompose $M$ into a Linear Combination of Unitaries (LCU), as it is done with $A^{-1}$ in \cite{QLSAchilds}, and how to simulate those linear operators efficiently. Notice that now $e^{-iM}$ is a unitary operator.

The first question is how to decompose $M$ in a linear combination of unitaries. If we choose, as it is frequently done, unitary operators of the form $e^{-iMt}$, we can write $M = \sum_i \beta_i e^{-iMt_i}$. Then, as both sides of the equality are diagonal in the same basis, one can consider this is equivalent to writing $x = \sum_i \beta_i e^{-ixt_i}$. The first idea is then to decompose $x$ as a Fourier series. Specifically, decomposing $x$ gives us
\begin{equation}
\begin{split}
    x &= \sum_{t=1}^\infty \frac{(-1)^t}{t} 2\sin(tx)\\
     &= i\sum_{t=1}^\infty  \frac{(-1)^t}{t} (e^{-itx}-e^{itx}).
\end{split}
\end{equation}
Since we cannot perform the entire summation, the question is how to truncate the series maintaining a given precision $\epsilon^{-1}$. Call $S_{\bar{t}}(x)$ the truncated series up to $\bar{t}$. Since all the singular values will be in $[1/\kappa_M,1]$, with $\kappa_M$ the condition number of $M$, then, we want to study the series in the range $[0,1]$. Notice that $f(x)=x$ is Lipschitz-continuous with constant $ 1$, so can use a result from \cite{jackson1930theory} to bound the error incur when truncating the series. According to Corollary I in \cite{jackson1930theory}, we can say that $|f(x)-S_{\bar{t}}(x)|\leq (a \log{\bar{t}})/\bar{t}$, where $a$ is a constant. Thus, if we want $|x-S_{\bar{t}}(x)|\leq\epsilon$, it is enough to choose $\bar{t}/\log \bar{t} = O(\epsilon^{-1})$. So the complexity is almost linear in $\epsilon^{-1}$, as we have to Hamiltonian-simulate $M$ during time $\bar{t}$. Furthermore, if we chose a Hamiltonian simulation algorithm, as $M$ is dense, the cost will be relatively large, $\Tilde{O}(\sqrt{n+m})$ if we were to use \cite{Dense_hamiltonian_simulation} for instance. In any case we would not obtain an exponential advantage, since Hamiltonian simulation of dense Hamiltonians is not expected to have complexity $\log n$ in general \cite{childs2009limitations}. Also notice that we cannot use the trick from the previous section, since that requires $M$ to be a density matrix and $\mathrm{tr} M = 0 \Rightarrow M$ is not a density matrix.

A second, better approach, also inspired by \cite{QLSAchilds}, is to decompose the matrix in Chebyshev polynomials, $x = \sum \gamma_i \mathcal{T}_{i}(x)$. In fact, this decomposition in polynomials is quite simple since the first Chebyshev polynomial is
\begin{equation}
    \mathcal{T}_{1}(x) = x,
\end{equation}
so the expansion will have a single term and will be exact, which is very positive. Let us now look at the cost of simulating a Chebyshev polynomial for a $\bar{n}\times \bar{n}$ $d$-sparse matrix $M$. One does that via Quantum Walks. Define a quantum walk as set of states $\{\ket{\psi_j}\in \mathbb{C}^{2\bar{n}}\otimes\mathbb{C}^{2\bar{n}}; j\in [\bar{n}]\}$ \cite{QLSAchilds}:
\begin{equation}
\begin{split}
    &\ket{\psi_j}:= \ket{j}\otimes \\
    &\frac{1}{d} \sum _{k\in [\bar{n}]; A_{jk}\neq 0}\left(\sqrt{A^*_{jk}}\ket{k}+\sqrt{1-|A_{jk}|}\ket{k+\bar{n}}\right).
\end{split}
\end{equation}
Then, one defines the isometry
\begin{equation}
    T= \sum_{j\in [\bar{n}]}\ket{\psi_j}\bra{j},
\end{equation}
and $S$ the Swap operator $S:\ket{j,k}\rightarrow\ket{k,j}$. The Walk operator is $W := S(2TT^\dagger - \mathbf{1})$. With these definitions, and using Lemma 15 of \cite{QLSAchilds} enunciated in the appendix \ref{sec:definitions}, one can prove that for any state $\ket{\psi}$, substituting $T$ by its corresponding unitary $T_U$ \cite{QLSAchilds}
\begin{equation}
\begin{split}
    &T_U^\dagger W T_U \ket{0^{\lceil \log 2\bar{n}\rceil +1}}\ket{\psi}\\& = \ket{0^{\lceil \log 2\bar{n}\rceil +1}} \mathcal{T}_{1}(\overline{M})\ket{\psi} + \ket{\Phi^{\perp}}\\
    &=\ket{0^{\lceil \log 2\bar{n}\rceil +1}} \overline{M}\ket{\psi} + \ket{\Phi^{\perp}},\label{Quantum walk}
\end{split}
\end{equation}
for $\overline{M}= M/d$, and $\ket{\Phi^{\perp}}$ such that $(\ket{0^{\lceil \log 2\bar{n}\rceil +1}}\bra{0^{\lceil \log 2\bar{n}\rceil +1}}\otimes \mathbf{1})\ket{\Phi^{\perp}}=0$. 

Let us first study the cost of applying $T$. According to Lemma 10 in \cite{berry2015hamiltonian}, to implement $T$ one must apply $\log d $ Hadamard gates, followed by $O(1)$ calls to an oracle $O_F$ that outputs the column index of the $l$-th non-zero element of a row $j$ of the matrix, $O_F:\ket{j,l}\rightarrow\ket{j, f(l,j)}$; and the sparse-access oracle $O_M$, that outputs entry $M_{jk}$ on input $(j,k)$. For simplicity and without loss of generality let us assume that in \eqref{new Matrix system} all entries of $A$ are non-zero. If such is the case, then the cost of the oracles is not high either. The first oracle outputs $O_F:\ket{j,l}\rightarrow\ket{j, l}$ or $O_F:\ket{j,l}\rightarrow\ket{j, l+n+1}$ depending on whether we are on the last $n+1$ rows or in the first $m+1$, respectively. The second oracle performs $O_M:\ket{j,k}\ket{z}\rightarrow\ket{j,k}\ket{z\oplus M_{jk}}$, which is nevertheless no more restrictive than the qRAM model that we were using to solve the quantum SVM.

What is the cost of implementing $W$ then? In the proof of their theorem 4, \cite{QLSAchilds} cites Lemma 10 from \cite{berry2015hamiltonian} to explicitly state that if we want to simulate $W$ with error $\epsilon'$, its complexity and the complexity of \eqref{Quantum walk} is $O(\log \bar{n} + \log^{2.5}(\kappa d / \epsilon'))$. This is not surprising since the cost only depends on $T$, its inverse $T^\dagger$, and a Swap gate. Notice that our matrix is dense, so $d = O (\bar{n})=O(n+m)$, but still the cost of this process is polylogarithmic in all variables! In fact this method seems applicable to any matrix $M$, sparse or not, whose entries are accessible through a sparse-access oracle in the form of a qRAM. One may even say that the qRAM is hiding the cost, since it requires time to prepare the data and also has $O(n)$ quantum gates even if it only uses a small number of them. Notice also that the linear dependence on $d$ and $\kappa$ for the Quantum Linear System Algorithm presented in \cite{QLSAchilds} does not come from implementing $W$ but because one has to apply it $O(d\kappa)$ times in the series expansion of $A^{-1}$. In our case we only have to apply $W$ once, so we do not have such linear complexity term.

Finally, we measure the amplitude of $\ket{1}$ of the ancilla that we attached to the state to mark the entry of $b$ after \eqref{second matrix system}. The amplitude of such state is, after taking into account a normalization factor for $M$ and $||(b,\vec{\alpha})||$, precisely $|\vec{w}_{n+1}|=\sqrt{\sum_{i=1}^m (\sum_{j=1}^{n+1} x_{ij,x_{n+1}} \alpha_{j,x_{n+1}})^2}$, what we were looking for. We can measure that amplitude using Amplitude Estimation.

In appendix \ref{sec:Norm_appendix} we explain in greater detail how to correct the value extracted with Amplitude Estimation, to find the norm $|\vec{w}_{n+1}|$ provided either the Linear Combination of Unitaries with the Fourier series, or the Chebyshev polynomial approach. The cost of Amplitude Estimation is $O(\epsilon^{-1})$.
Having calculated $|P_c(\vec{x}_{n+1})|$ and $|\vec{w}_{n+1}|$, the informativeness of a given point is just the product of both.

\subsection{\label{sec:Target} Finding the target point $\vec{x}_{n+1}$}

Now that we know how to calculate the `informativeness' of a given point, let us explain why the Algorithm \ref{Main algorithm} finds, with probability $1-e^{-\beta}$ a point that is in the quantile $1-1/C$ of informativeness. The arguments presented here are the same to those in appendix \ref{sec:classical_strategies}. 

If we sample uniformly at random, and calculate the informativeness of $c$ points, the probability that none of those points is in the quantile $1-1/C$ is clearly 
\begin{equation}
    p = \left(1-\frac{1}{C}\right)^c.
\end{equation}
We want to make such probability $p< e^{-\beta}$, so that at least one is in the $1-1/C$ quantile. That means
\begin{equation}
    c > \frac{-\beta}{\log \left(1-1\frac{1}{C}\right)}.
\end{equation}
Since 
\begin{equation}
    \lim_{C\rightarrow\infty} \frac{\frac{-1}{\log \left(1-\frac{1}{C}\right)}}{C} = 1,
\end{equation}
it is clear then that the number of sampled points should be $c = O(\beta C)$. This explains why our procedure is efficient in these variables. More information on alternative strategies can be found in the appendix \ref{sec:classical_strategies}.

\section{\label{sec:results}Main results}

In this article we propose a theoretical framework that allows us to think of active learning as sampling the most promising new points to be classified, so that the minimum distance between classes can be found, and theorem \ref{Teorema 5} used. We also propose a quantum algorithm that would allow us to perform the sampling efficiently with a exponential advantage over what is achieved classically.

In the quantum algorithm we do not use neural networks but rather a SVM \cite{SVM}, although it might be possible to use the general strategy in other setups using quantum neural networks.

Let us explain further the complexity comparison with the classical algorithms. One may argue that a good quantum strategy to solve Problem \ref{Informative sampling problem} would be to use Amplitude Estimation and bisection search to find a threshold for the quantile $1-1/C$, for example the $1\%$ most relevant points, in which case $C=100$. Then, one may use Amplitude Amplification to find one of the points in such quantile. This achieves an exponential advantage over the classical case when one wants to find such points using classical computing with certainty. In the usual case of Amplitude Amplification and Amplitude Estimation we have an oracle that tells us whether an element is marked or not, and this yields a quadratic advantage with respect to the classical case. However, in our situation, being a good point or not depends on its relative `informativeness' with respect to other points. This means that classically, in order to assert with certainty that a given point is within the top $1/C=1\%$ quantile, one should first calculate the informativeness of $0.99N$ points, out of $N$ points. This is clearly prohibitive, since $N = O(l^m)$, where $m$ is the dimension of the space, and $l$ its discretisation. Notice for example that for a $n_0\times n_0$ image, the dimension of the image would be $m = 3 n_0^2$, due to the three colours or channels needed to define each pixel. 

In contrast, if we want to solve this problem using Amplitude Estimation we do not incur in such cost. What we do is find, using bisection and Amplitude Estimation, an informativeness threshold above which there are only $1/C$ of the most informative points. Once that is the case, we can mark those points and use Amplitude Amplification to find them \cite{Amplitude_estimation}. The cost will then be $O(C\epsilon^{-1}\beta)$ for a given precision $\epsilon$ in the threshold, and success probability $1-e^{-\beta}$, and crucially $C$ independent of $N$. Amplitude Amplification also bears a cost $O(\sqrt{C})$.

On the other hand, one can also find probabilistic classical strategies that output a point that is in the $1-1/C$ quantile with probability exponentially high, $1-e^{-\beta}$, and the complexity would be polynomial in parameters $C$ and $\beta$. Since our quantum strategy would also have some exponentially small probability of failure, and the complexity would also be polynomial in those parameters, the advantage would be unclear. We discuss this point in depth in the appendix \ref{sec:classical_strategies}, for a problem that generalises our own, and find that this advantage would be quadratic in some parameters.

However, the algorithm that we present here achieves an exponential advantage over its classical counterpart. The speedup is due to a faster calculation of the `informativeness' of a given point, thanks to the use of qRAMs, the algorithm for quantum SVM \cite{QSVM}, and an approach to perform matrix-vector products using Chebyshev polynomials taking advantage of techniques developed in \cite{berry2015hamiltonian} and \cite{QLSAchilds}; in conclusion, a better use of quantum linear algebra techniques. Our algorithm here has overall complexity $\Tilde{O}(C\epsilon^{-1}\beta(o||F||^2\kappa_{\mathrm{eff}}^{-3}\epsilon^{-3}+ \mathrm{poly}\log (\kappa_M\epsilon^{-1}(n+m))))$, where $||F||$ is the Frobenius norm of the matrix in \eqref{system of equations}, $o$ the order of the kernel, $\kappa_{\mathrm{eff}}=O(1)$ an effective condition number that we choose, $\epsilon$ the precision on the `informativeness' of a given point, $1-e^{-\beta}$ the success probability with which one wants to be sure that the chosen point is in the $1-1/C$ quantile, and $\kappa_M$ the condition number of matrix M appearing in \eqref{new Matrix system}. 

In contrast, a classical algorithm would have polynomial complexity in $n$ and $m$ due for instance to matrix-vector product \eqref{new Matrix system}, or the final calculation of the norm of $|\vec{w}|$, central to our discussion.

In the next subsection we lay out the general strategy of our paper to solve this problem.

\section{\label{sec:Complexity} Complexity}

In this section we want to give a calculation of the complexity of the proposed quantum algorithm. 

The first step is calculating the probability that a given point of the high dimensional space is in a certain class, $P_c(\vec{x}_{n+1})$. This implies solving a quantum SVM \cite{QSVM}, with complexity $O(o ||F||^2\epsilon^{-3} \kappa_{\mathrm{eff}}^{-3} \log (mn))$, where $o$ is the order of the kernel, $||F||$ is the Frobenius norm of the matrix that appears in \eqref{system of equations}, $m$ the dimension of the space, $n$ the number of already classified points, and $\kappa_{\mathrm{eff}}$ is an effective condition number which can be taken $O(1)$ for this problem. 

Reading out the probability using Amplitude Estimation in the Swap Test costs an additional multiplicative $O(\epsilon^{-1})$, so the overall complexity of this step is $O(o ||F||^2 \epsilon^{-4} \kappa_{\mathrm{eff}}^{-3} \log (mn))$.  Since we are assuming that we have scaled the matrix to obtain the largest eigenvalue equal to 1, the procedure of \cite{QSVM} implies that the smallest one that we take into consideration is $\lambda^{-1} = \kappa_{\mathrm{eff}}$. The low-rank approximation induces an additional error of $O(n^{-1/2})$, as it is explained in the supplementary material of the original article \cite{QSVM}. 

Next step is calculating the Quantum SVM for the system with the added point, again with complexity $O(o ||F||^2 \epsilon^{-3} \kappa_{\mathrm{eff}}^{-3} \log (mn))$.
Then, preparing $\ket{\vec{w}_{n+1}}$ has complexity $O(\text{poly}\log (\kappa_M \epsilon^{-1}(n+m)))$, as reported in the corresponding section. Calculating $|\vec{w}_{n+1}|$ also have an additional multiplicative complexity of $\Tilde{O}(\epsilon^{-1})$ due to the Amplitude Estimation, given the procedure in appendix \ref{sec:Norm_appendix}.

Finally, if we are interested in finding a point in the $1-1/C$ quantile of `informativeness' with probability $1-e^{-\beta}$, then the procedure explained in section \ref{sec:Target} implies iterating the procedure over $O(C\beta)$ points and selecting the best.

Thus, in general, the complexity will be $\Tilde{O}(C\beta\epsilon^{-1}(o||F||^2 \kappa_{\mathrm{eff}}^{-3}\epsilon^{-3}+ \text{poly}\log (\kappa_M\epsilon^{-1}(n+m))))$, with error $\Tilde{O}(\epsilon + n^{-1/2})$. The $n^{-1/2}$ comes from the low-rank approximation of \cite{QSVM}, as it is explained on its appendix C.

\section{\label{sec:Conclusion}Conclusion}
In the previous sections we have seen that provided the use of qRAMs, it could be possible to establish a polynomial-complexity sampling procedure which lets us know what are the most promising points to be added to the training set in order to get a better approximation of the minimum distance between classes, $r_p$. Recall that knowing $r_p$ is a requisite to applying a $\delta$-cover and, using theorem \ref{Teorema 5}, avoid adversarial examples. This protocol is heuristic, which means that in order to check its actual performance we have to run it in a realistic quantum computer.

In this paper we have presented a procedure that allows to solve this problem efficiently, in time polylogarithmic in the dimension of the space $m$ and number of already classified points, $n$; and polynomial in $C$, $\beta$ and the precision $\epsilon^{-1}$ of the informativeness estimate.

There are nevertheless two important shortcomings of our algorithm. The first is that, since we rely heavily on reference \cite{QSVM}, and they use qRAMs to achieve their speedup, we also need qRAMs. The second is that there are some minor parts of the general algorithm that will necessarily have polynomial complexity in the dimension of the space $m$. Those are: sampling initial points, and loading their value in the qRAM. We strongly believe that this should nevertheless be no problem. The sampling procedure for instance will hardly be inefficient and can be easily done while the previous point is being processed. The loading in the qRAM is perhaps more expensive, but may be done while the point is asked to be classified by a human. Therefore we believe that the complexity result is still very valid in practice.

Finally, we want to recall that our work relies heavily on reference \cite{QSVM}. At the time that article was written, there was no known method to solve low-rank linear systems of equations in polylogarithmic time. However, \cite{chia2018quantum,arrazola2019quantum} recently proposed a method that achieves precisely that, complexity $O(||F||^6 k^6 \kappa^{16} \epsilon^{-6})$, $k$ the rank. It relies on the techniques developed by Ewin Tang \cite{ewin-tang}, which found a classical algorithm taking inspiration from \cite{Q_recommendation_sys}. Therefore, one could also solve the linear system of equations that appear in this case in polylogarithmic time in the dimension $n$. In the published version of this article it is argued that the algorithm could probably easily be dequantized, following such techniques, since the most complicated part is the solution of the system of equations. Here, we have added appendix \ref{sec:dequantization} to explain how to do it.

In conclusion, in this article we have framed a possible solution to adversarial examples using active learning, and a sampling methodology to find the most important points that could be added to the training set. We have also presented an algorithm that provides an exponential advantage over its classical counterpart, provided the use of qRAMs.

\section*{Acknowledgements}
We would like to thank Santiago Varona for useful comments on the manuscript, as well to Jaime Sevilla, Nikolas Bernaola and Javier Prieto for pointing us to useful statistic results for Appendix \ref{sec:classical_strategies}.
We acknowledge financial support from the Spanish MINECO grants MINECO/FEDER Projects FIS 2017-91460-EXP, PGC2018-099169-B-I00 FIS-2018 and from CAM/FEDER Project No. S2018/TCS-4342 (QUITEMAD-CM). The research of M.A.M.-D. has been partially supported by the U.S. Army Research Office through Grant No.  W911NF-14-1-0103. P. A. M. C. thanks the support of a FPU MECD Grant. 

\bibliographystyle{ieeetr}
\bibliography{bibliography}

\appendix

\section{\label{Provably_robust} A provably robust classifier}

In this appendix we would like to review the theorem of \cite{Geometry_of_adversarial_examples} that is the basis for the provably robust classifier. We introduce informal definitions for a $\delta$\textit{-cover} of a manifold, the $\epsilon_0$\textit{-neighbour} of a submanifold, and finally what is an \textit{adversarial example}. A formal definition for those concepts can be seen at appendix \ref{sec:definitions}.

Informally stated, a $\delta$\textit{-cover} of a manifold is a set of points $\{\vec{x}_i\}$ of the manifold such that the set of balls with centers $\{\vec{x}_i\}$ and radious $\delta$ contains the manifold. Or in other words, any point at the manifold is at most $\delta$-far from a point from $\{\vec{x}_i\}$. Relatedly, an $\epsilon_0$\textit{-neighbour} of a given submanifold is composed of all those points in the manifold at most $\epsilon_0$-far from the submanifold. In both cases we are assuming a $p$-norm.

\begin{figure}
    \centering
    \includegraphics[scale=.35]{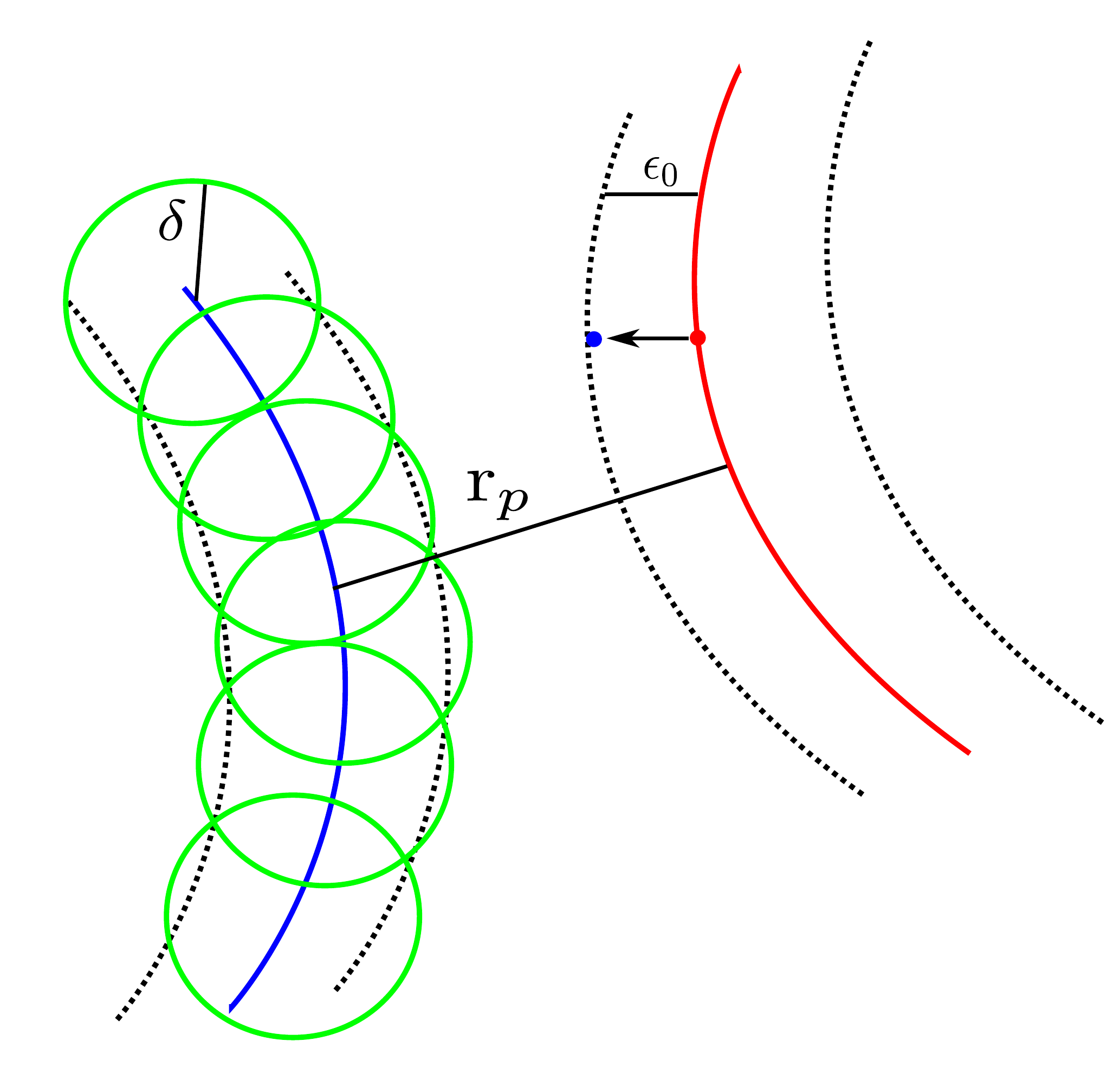}
    \caption{Submanifold $\mathcal{M}$ made of two classes separated by a distance $r_p$. The space near each class is $\mathcal{M}^{\epsilon_0}$. We also represent a $\delta$-cover of the blue class. The red point is in the red class, but when $\epsilon_0$-shifted, can be mistakenly classified as blue. It would be an adversarial example that we are trying to avoid.}
    \label{fig:neighbour}
\end{figure}

An example of $\delta$-cover is depicted in \ref{fig:neighbour} in green and will be a key ingredient to avoid adversarial examples. With perhaps an abuse of notation, we will call a $\delta$-cover simultaneously to the set of $\vec{x}_i$ points that are in the center of these balls, and to the balls themselves. This means that for $\delta$ the parameter that controls how coarse or fine is the sampling, a $\delta$-cover is a coarse-grained sampling of each class. Following our previous example, an example of a $\delta$-cover is a set of images of cars, for example, such that any possible image of a car is no further than $\delta$-far to one of the training set. Notice that one can measure distance between images by the distance between the vectors containing the amount of green, blue and red of each pixel, in a $p$-norm. 

For the next definition we will need the notion of a \textit{classifier}, a function $f$ that, given a point $x$ is able to predict a label $y$. An $\epsilon_0$-neighbour is also depicted in figure \ref{fig:neighbour} in dashed lines around the red and blue submanifolds. Here, $\epsilon_0$ has the meaning of the robustness against perturbations. For example, take an image of a car and its corresponding vector. $\epsilon_0$ is the amount one can perturb the vector without fooling the classifier. Then, $\mathcal{M}^{\epsilon_0}$ is the space of such perturbations of size $\epsilon_0$, for each class within submanifold $\mathcal{M}$, that contains the different classes. 

 Provided the previous definitions, an $\epsilon_0$\textit{-adversarial example} can be described as a point in the manifold that, being $\epsilon_0$-close to a class or submanifold, is mistakenly classified as an example of another class. For such definition to make sense $\epsilon_0$ must be smaller than the distance between two classes $r_p$. Else, a given point in the region $\mathcal{M}^{\epsilon_0}$ could be $\epsilon_0$-close to two classes at the same time.
An example of an adversarial example is pictured in figure \ref{fig:neighbour}. The blue point is $\epsilon_0$-close to the red class but classified as blue.

The intuition of \cite{Geometry_of_adversarial_examples} to avoid $\epsilon_0$-adversarial examples is to cover all classes with a $\delta$-cover, such that any point in the $\epsilon_0$-neighbour of the classes is $\delta$-close to a correctly classified point. What we want to find out is how big can $\delta$ be in order to maintain protection against adversarial examples. Such $\delta$ will depend crucially on the minimum distance of separation between classes, $\delta < r_p -\epsilon_0$, and thus our main objective in this article is to sample efficiently in an active learning setting, to find this minimum distance $r_p$. The precise statement of the theorem can be found in appendix \ref{sec:definitions}, but
intuitively we are trying to cover the space of $\mathcal{M}^{\epsilon_0}$ corresponding to each class, with sets of balls, such that the sets do not overlap but the balls are as large as possible to avoid sampling more than it is needed.

Therefore, if we knew the minimum distance of separation of two classes, $r_p$, we would be able to produce a cover of the two classes that avoids the adversarial examples. However, finding $r_p$ is not easy, because we only have an upper bound to the minimum distance of samples between two classes. The generalisation to a small arbitrary number of classes is usually done by pairs. The problem we aim to solve in this paper is finding this minimum distance between the two classes $r_p$, because if we overestimate $\delta$ we would not be able to use theorem \ref{Teorema 5}, and if we underestimate it, the $\delta$-cover would be more expensive to establish.

Given these definitions and the previous theorem, we can think of this as a procedure to establish an $\epsilon_0$-tolerance to perturbations. The robustness provided by the $\delta$-cover qualitatively means that, if we provide a cover of the class with balls of size $\delta$, then any point $\epsilon_0$-near the class, in $\mathcal{M}^{\epsilon_0}$, will be correctly classified. Thus, our classifier will be robust to perturbations. Therefore, the level of robustness $\epsilon_0$ is something we choose, and $r_p$ is the unknown we are looking for that would allow us to calculate $\delta$.

\section{\label{sec:definitions} Technical definitions and previous results}

In the main text we gave informal definitions about concepts needed to understand theorem \ref{Teorema 5}. Here we give them rigorously. The first concept we need to introduce is that of a $\delta$-cover of a manifold, which will be used in that theorem.

\begin{Definition}
Given a manifold $\mathcal{M}$, a $\delta$-cover of such manifold is a set of balls of radius $\delta$, $\bigcup_{i\in I} B(x_i,\delta)$, such that $\forall x \in \mathcal{M}$ $\exists j\in I | x\in B(x_i,\delta)$.
\end{Definition}

Additionally, in order to understand the meaning of an adversarial example, we need to define an $\epsilon$-neighbour:

\begin{Definition}
Given a manifold $\mathcal{M}\subset \mathbb{R}^m$, an $\epsilon$-neighbour of $\mathcal{M}$ in the norm $p$, $\mathcal{M}^\epsilon$, is the set of points $x\in\mathbb{R}^m$ such that the $p$-distance of $x$ to $\mathcal{M}$, $d_p(x,\mathcal{M})\leq\epsilon$.
\end{Definition}

Finally, an adversarial example is defined as

\begin{Definition}
Let $\mathcal{M}\subset \mathbb{R}^m$ be a manifold containing several disjoint parts called classes $C_i$ separated by a distance $r_p\gg \epsilon$ in the norm $p$, and a classifier $f$ reasonably well trained to distinguish between those classes. An $\epsilon$-adversarial example of such classifier in the norm $p$ is a point $x$ whose $p$-distance to a given class $C_0$ is smaller than $\epsilon$, but it is classified to be in class $C_1$. That is $d(x,C_0)\leq \epsilon$ and $f(x)=C_1$.
\end{Definition}

With all the previous definitions, we can state the theorem \ref{Teorema 5} that guarantees resistance to adversarial examples.

\begin{Theorem}\label{Teorema 5} \textbf{(Khoury and Hadfield-Menell)} \cite{Geometry_of_adversarial_examples} 
Let $\mathcal{M}\subset \mathbb{R}^m$ be a $k-$dimensional manifold that contains each of the classes, and let $ r_p$ be the minimum separation distance between two classes in norm $p$, $r_p>\epsilon$. Let $\mathcal{L}$ be a learning algorithm, and $f_{\mathcal{L}}$ the classifier it produces. Assume that for any point $x$ in the training set $X_{\mathcal{L}}$ with label $y$, and any point $\hat{x}\in B(x,r_p)$, the learning algorithm classifies $f_\mathcal{L}(\hat{x})= f_\mathcal{L}(x)=y$. 

We then have the following guarantee:
If $X_{\mathcal{L}}$ is a $\delta$-cover for $\delta < r_p-\epsilon$ then $f_{\mathcal{L}}$ correctly classifies $\mathcal{M}^{\epsilon}$, that is, an $\epsilon$-neighbour of $\mathcal{M}$. 
\end{Theorem}

In this appendix we also give an overview of the quantum-accessible data structure that we use to perform the Hamiltonian Simulation of Dense matrices

\begin{Theorem} \cite{kerenidis2017recommendation, block-encoding_1}:
Let $M\in \mathbb{R}^{n'\times n'}$ be a matrix. Let $w$ be the number of nonzero entries. Then there exists a quantum-accessible data structure of size $O(w\log^2(n'^2))$, which takes time $O(\log (n'^2))$ to store or update a single entry. 
Using this data structure, there is a quantum algorithm able to perform the following maps with error $\epsilon$:
\begin{equation}
    U_\mathcal{M}:\ket{i}\ket{0}\rightarrow \frac{1}{||M_{i\cdot}||}\sum_{j}M_{ij}\ket{ij};
    \label{mathcal M}
\end{equation}
\begin{equation}
    U_\mathcal{N}:\ket{0}\ket{j}\rightarrow \frac{1}{||M||_F}\sum_i ||M_{i \cdot}|| \ket{ij};
    \label{mathcal N}
\end{equation}
where $||M_{i \cdot}||$ is the $l_2-$norm of row $i$ of $M$. The complexity of this algorithm is $O(\text{poly}\log(n'/\epsilon))$.
This means in particular that given a vector $f$ in this data structure, we can prepare an $\epsilon$ approximation of it, $1/||v||_2 \sum_i v_i \ket{i}$, in time $O(\text{poly}\log(n'/\epsilon))$.
\end{Theorem}
 \qed

The previous data structure can be used for dense Hamiltonian simulation \cite{Dense_hamiltonian_simulation}, into the main algorithm of \cite{QLSAchilds}

\begin{Theorem}\cite{Dense_hamiltonian_simulation} \textbf{Hamiltonian simulation of Dense Hamiltonians}.
Let $H$ be a $2^n$-dimensional Hamiltonian stored in the quantum-accessible data structure. Then there is a quantum algorithm able to simulate it in complexity
\begin{equation}
    O\left(t\Lambda n \log^{5/2}(t\Lambda \epsilon^{-1}) \frac{\log (t||H|| \epsilon^{-1})}{\log \log (t||H|| \epsilon^{-1})}\right),
\end{equation}
for $t$ the simulation time, $\epsilon^{-1}$ the precision and $\Lambda = \max\{||H||,||H||_1 \}$, taking into account that $||H||_1\leq \sqrt{2^n}||H||$, but taking $||H||$, the spectral norm, as measure of cost.

\end{Theorem}

The Hamiltonian simulation is very useful to solve linear systems of equations. The first and most famous of those algorithms is usually called HHL after their discoverers Harrow, Hassidim and Lloyd.

\begin{Theorem} \cite{HHL} \textbf{(HHL)}
Let $M$ be an $n' \times n'$ Hermitian matrix (if the matrix is not Hermitian it can be included as a submatrix of a Hermitian one) with condition number $\kappa$ and $M$ having an sparsity $d$ (at most $d$ nonzero entries in each row).

Let $b$ be an $n'$-dimensional unit vector, and assume that there is an oracle $\mathcal{P}_b$ which produces the state $\ket{b}$, and another $\mathcal{P}_M$ which, taking $(r, i)$ as input, outputs the location and value of the i’th nonzero entry in row $r$ of $M$. Let
\begin{equation}
    x=M^{-1}b, \qquad \ket{x}=\frac{x}{||x||}.
\end{equation}

Then, there is an algorithm that outputs $\ket{x}$ in complexity $O(d \kappa^2 \epsilon^{-1} \text{poly}\log n')$.

\end{Theorem} \qed

However, since it was published, more efficient methods have been published. In particular, \cite{Ambainis} and \cite{QLSAchilds} achieved improvements in the condition number and precision respectively. We present here the main result from the later.

\begin{Theorem}\cite{QLSAchilds}
Let $M$ be an $n' \times n'$ Hermitian matrix (if the matrix is not Hermitian it can be included as a submatrix of a Hermitian one) with condition number $\kappa$ and $M$ having an sparsity $d$ (at most $d$ nonzero entries in each row).

Let $b$ be an $n'$-dimensional unit vector, and assume that there is an oracle $\mathcal{P}_b$ which produces the state $\ket{b}$, and another $\mathcal{P}_M$ which, taking $(r, i)$ as input, outputs the location and value of the i’th nonzero entry in row $r$ of $M$. Let
\begin{equation}
    x=M^{-1}b, \qquad \ket{x}=\frac{x}{||x||}.
\end{equation}
Then, \cite{QLSAchilds} construct an algorithm relying on Hamiltonian simulation that outputs the state $\ket{x}$ up to precision $\epsilon$, with constant probability of failure (i.e., independent from the problem parameters), and makes
\begin{equation}
    O(d\kappa\log^{2.5}(\kappa/\epsilon))
\end{equation}
uses of $\mathcal{P}_M$ and 
\begin{equation}
    O(\kappa \sqrt{\log(\kappa/\epsilon)})
\end{equation}
of $\mathcal{P}_f$; and has overall time complexity 
\begin{equation}
    O(d\kappa \text{ poly}(\log(n'd\kappa/\epsilon))).
\end{equation} \qed
\end{Theorem}

Also in the same article, \cite{QLSAchilds}, the authors present the Lemma 15 and 16, which we use
\begin{Lemma} \cite{QLSAchilds} \label{Lemma 15}
Let $\ket{\lambda}$ be an eigenvector of $H = M/d$, $M$ a $\bar{n}\times \bar{n}$ matrix with sparsity $d$. Let $\lambda\in (-1,+1)$ the corresponding eigenvector. Within the invariant subspace $\{T\ket{\lambda}, ST\ket{\lambda}\}$, the walk operator $W$ has block form \begin{equation}
    \begin{pmatrix}
    \lambda & -\sqrt{1-\lambda^2}\\
    \sqrt{1-\lambda^2} & \lambda
    \end{pmatrix}.
\end{equation}
In particular, this will have the consequence that 
\begin{equation}
    W^j T\ket{\lambda} = \mathcal{T}_j(\lambda)T\ket{\lambda}+ \sqrt{1-\lambda^2}\mathcal{U}_{j-1}(\lambda)\ket{\perp_\lambda},
\end{equation}

with $\mathcal{T}_j$ is the $j$th Chebyshev polynomial of the first kind, and $\mathcal{U}_j$ of the second kind; and $\ket{\perp_\lambda}$ a state in $\{T\ket{\lambda}, ST\ket{\lambda}\}$ perpendicular to $T\ket{\lambda}$. 
\end{Lemma}
Notice that this lemma is the origin of \eqref{Quantum walk}.

 Finally, let us state the theorem of Amplitude Estimation, which we use through the text.

\begin{Theorem} 
\textbf{(Amplitude Estimation)} \cite{Amplitude_estimation}: For any positive integer $k$, the algorithm Amplitude Estimation outputs an estimate $0\leq\Tilde{a}\leq 1$ of the desired amplitude $a$ such that
\begin{equation}
    |a-\Tilde{a}|\leq 2\pi k \frac{\sqrt{a(1-a)}}{J}+k^2\frac{\pi^2}{J^2},
\end{equation}
with success probability at least $\frac{8}{\pi^2}$ for $k=1$, and with success probability greater than $1-\frac{1}{2(k-1)}$ for $k\geq 2$. $J$ is defined as the number of times we need the implementations of the oracle that tells whether an element is marked, for Amplitude Estimation. Also, if $a=0$ then $\Tilde{a}=0$, and if $a=1$ and $J$ even, then $\Tilde{a}=1$. 
\end{Theorem}  \qed

\section{\label{sec:classical_strategies} Probabilistic classical strategies to sample with certainties.}

In the main text we have focused on the particular problem of sampling the most relevant points to determine the distance between different classes in the context of classifying. Here we focus on a more general setting, where one is given a oracle scoring function $s$ and wants to sample the points of the space with higher score, with exponentially high probability. We will see that some quadratic advantage is possible for non-deterministic functions $s$.

Imagine we have a $m$-dimensional space $\mathcal{S}$, which we discretise in $n_0$ points in one dimension such that in the end it contains $N = n_0^m$ points. This is for example what happens when we have the space of images of $n_0 = (n_p \times n_p)$ pixels, each displaying $256$ possible values. If the image is in colour then to each pixels three such values ranging from $0$ to $255$ will be assigned indicating the coordinates (red, green, blue), so that the number of possible points or images in the space would be $N = (256)^{3n_0}=(256)^{6n_p}$.  

Another example could be, in the setting of reinforcement learning, the range of policies parametrised by $n$ parameters, each of which can again take a range of discrete values. The policy of an agent is a function that takes as input the state of the agent or the observation it makes, and outputs an action which seeks to maximize the reward the agent gets. In this case the space $\mathcal{S}$ would be the space of possible policies. As we can see in both cases the number of points in the space is exponential in the dimension $d$.

Suppose that we also have some kind of scoring function, which we shall call $s: \mathcal{S}\rightarrow \mathbb{R}$, which may or may not be deterministic, and is given to us as an oracle. The problem we aim to solve is finding a point which is in the top $1/C$ fraction of points attaining the highest score, with some exponentially high probability. That it, we want to find one point $x_0$ such that $P(s(x_0) \text{ in top }1/C)=1-p $ with $p= e^{-\beta}$. Thus, $C$ and $\beta$ are the parameters we want to vary. 

If the function $s$ is probabilistic, it will output a score when evaluated on point $x$ that follows a probability distribution with average $\mu_x$ and variance $\sigma$, $\mathcal{D}(\mu_x,\sigma)$. In this case the objective is to achieve the same guarantees for the average value of the evaluation of $x_0$. In other words, $P(\mu_{x_0} \text{ in top }1/C)=1-p $ with $p= e^{-\beta}$. Let us summarise then the problem.

\begin{Problem}
\textbf{(Sampling with statistical guarantees)}. Let $\mathcal{S}$ be a $m$-dimensional discrete space. Let $s: \mathcal{S}\rightarrow \mathbb{R}$ be a scoring function that is given to us as an oracle. The problem is to return a point $x$ that, with probability $1-e^{-\beta}$ is in the percentile $p_C = 1-1/C$ of $s$. If $s$ is non-deterministic, but rather $s(x)\in \mathcal{D}(\mu = \mu_x,\sigma^2)$, $\mathcal{D}$ a probability distribution, then return a point $x$ such that $\mu_x$ is in the previously mentioned percentile.
\end{Problem}

In the main text we initially compared our strategy with the most Naive classical one: if we want certainty that the point we choose really is in the top $1/C$ that means having to sample $O(1/N)$ of them. However we could argue that an exponentially small probability of the point not being in the top $1/C$ of the most interesting ones is in fact a better comparison against our algorithm. After all, our algorithm does also achieve exponentially small probability of choosing the wrong point. 

This appendix aims to propose a quantum strategy that allows us to efficiently sample one such point even if variance $\sigma$ is large when compared against the range of values that $\mu_x$ may take, and compare it against classical strategies. We will see that the quantum strategy obtains a quadratic speedup on some parameters, due to the use of Amplitude Estimation \cite{Amplitude_estimation} and Amplitude Estimation. 

Let us start with the most naive of the classical strategies. If we try to sample points from the space until we get certainty that the best point of those sampled is one of the seeked ones, the cost will be exponential even in the deterministic case, taking $O(N(1-1/C))$ evaluations of $s$.

But if we only want statistical guarantees then one can do much better. The first of the probabilistic classical strategies can be called `greedy'.

\subsection{\label{sec:Greedy_strategy} The classical greedy strategy}

Suppose that we are in the deterministic case. The greedy strategy consists on following process:
\begin{enumerate}
    \item Uniformly at random, sample a single point $x$ from the entire space $\mathcal{S}$ of possible candidate points.
    \item Evaluate $s(x)$.
    \item If $s(x)$ is higher than any of the previously evaluated ones, substitute the previous maximum point by $x,s(x)$.
\end{enumerate}

By hypothesis we want to get a point that is in the top $1/C$, what means that each time we sample a point, we have a $1/C$ probability of spotting one. As this process gets the best of $c$ candidates, the probability that the best one is not in the top $1/C$ decreases as $(1-1/C)^c$. Complexitywise, this means that we want to find $c$ such that 
\begin{equation}
    p > \left(1-\frac{1}{C}\right)^c,
\end{equation}
with $p$ the probability of outputting a wrong point. Thus
\begin{equation}
    c > \frac{\log p}{\log \left(1-\frac{1}{C}\right)}.
\end{equation}

Obviously both the nominator and the denominator are negative. Let us analyse this complexity by parts. First notice that 
\begin{equation}
    \lim_{C\rightarrow\infty} \frac{\frac{-1}{\log \left(1-\frac{1}{C}\right)}}{C} = 1,
    \label{complexity C}
\end{equation}
what means that $\frac{-1}{\log \left(1-\frac{1}{C}\right)} = O(C)$, matching the complexity of our algorithm for the $C$ variable. On the other hand, as we make $p \rightarrow 0$, $-\log p$ goes very quickly to infinity. However if we make $p$ exponentially small, $p = e^{-\beta}$, then $-\log p = -\log e^{-\beta} = \beta$. And this would imply that $c= O(\beta C)$. 

Let us now consider the case where $s$ is not deterministic but rather follows a distribution $\mathcal{D}(\mu_x,\sigma)$. In such case the greedy procedure becomes more complicated. 

Let us introduce some notation. Assume that we name the points $x_0,...,x_c$ such that for the first evaluation $s(x_0)\geq ...\geq s(x_c)$. Also denote by $m_i$ the number of times $x_i$ has been evaluated, since we will need to evaluate several times the same $x_i$.

In order to analyse the case for non-deterministic function $s$, we will make use of the Central Limit Theorem:
\begin{Theorem}
Central Limit theorem. Let $X$ be a random variable that follows some probability distribution with average $\mu$ and variance $\sigma^2 < \infty$. If we sample $n$ times independently from such probability distribution, the sample average $\hat{\mu}$ will approximately follow, for $n$ large, a distribution with average $\mu$ and variance $\sigma^2/n$.
\end{Theorem}

Now, suppose that we have called the oracle function $s$ $m_0$ times for $x_0$, and $m_1$ times for $x_1$. That means that the averages of the distributions, $\hat{\mu}_0$ and $\hat{\mu}_1$, will follow distributions with averages $\mu_0$ and $\mu_1$ and variances $\sigma^2/m_0$ and $\sigma^2/m_1$. Notice that the probability distribution of the difference of two normal distributions is again a normal distribution with mean $\mu_0-\mu_1$ and variance $\sigma^2/m_0 + \sigma^2/m_1$ \cite{difference_normals}. Thus, the probability that in fact $\mu_0 \geq \mu_1$ is given by
\begin{equation}
    \int_0^\infty \mathcal{N}(\mu_0-\mu_1, \sigma^2/m_0 + \sigma^2/m_1),
\end{equation}
$\mathcal{N}$ the normal distribution. The previous expression is equal to 
\begin{equation}
    \left[\frac{1}{2}+\frac{1}{2}erf\left(\frac{z-(\mu_0-\mu_1)}{\sqrt{\sigma^2/m_0 + \sigma^2/m_1}}\right)\right]_0^\infty,
\end{equation}
with $z$ the variable and $erf$ the error function. Simplifying, as the error function is antisymmetric,
\begin{equation}
   P_1= \frac{1}{2}+\frac{1}{2}erf\left(\frac{(\mu_0-\mu_1)}{\sqrt{\sigma^2/m_0 + \sigma^2/m_1}}\right). \label{Probability mu0 > mu1}
\end{equation}
One would like this to be larger than $1-p'$ such that $1-p\leq (1-p')^c$ and $1-p = 1-e^{-\beta}$, so 
\begin{equation}
    \frac{1}{2}+\frac{1}{2}erf\left(\frac{(\mu_0-\mu_1)}{\sqrt{\sigma^2/m_0 + \sigma^2/m_1}}\right)\geq 1-p'
\end{equation}
implies
\begin{equation}
    erf\left(\frac{(\mu_0-\mu_1)}{\sqrt{\sigma^2/m_0 + \sigma^2/m_1}}\right) \geq 1-2p',
\end{equation}
and 
\begin{equation}
    \sigma^2/m_0 + \sigma^2/m_1\leq \left(\frac{(\mu_0-\mu_1)}{erf^{-1}(1-2p')}\right)^2.
\end{equation}
Notice that when $p\rightarrow 0$, the denominator goes to infinity. Since $p'\geq 1-(1-p)^{1/c}$, that means that $erf^{-1}(1-2p') = erf^{-1}(-1+2(1-p)^{1/c})$. Then, if we take $c = O(C\beta)$
\begin{equation}
\begin{split}
    &\lim_{\beta\rightarrow\infty}\frac{erf^{-1}(-1+2(1-e^{-\beta})^{1/c})}{\sqrt{\beta}}=\\
    & \lim_{\beta\rightarrow\infty}\frac{erf^{-1}(-1+2(1-e^{-\beta}/c))}{\sqrt{\beta} }=1.
\end{split}
\end{equation}
Notice however that the above expression does not depend on $c$. If $m_0 = m_1$ then $erf^{-1}(1-2p') =O(\sqrt{\beta})$ and
\begin{equation}
    m_1 = m_0 \geq \frac{2\sigma^2}{(\mu_0-\mu_1)^2} O(\beta).
\end{equation}
Notice that we have to check this for every pair $(m_0, m_i)$ for all $i\in(1,...,c)$, and as we had said that $c = O(\beta C)$, the overall complexity is upper bounded by $\Omega(\beta^2 C)$. This result is consistent even if $m_0 \neq m_1$, as can be easily checked. In contrast, we shall see that using quantum procedures we can achieve a complexity $O(\beta C)$. 

Another interesting point to extract from the previous equation is that the complexity is greatly reduced whenever $\sigma^2\ll (\mu_0-\mu_1)^2$. In particular, when $\sigma^2 =0$ we recover the limit for the oracle function $s$ being deterministic.

But one may argue that actually one may be able to generalise the strategy. With this we mean that the objective is checking that $\mu_0$ is in the top $1/C$, not that it is better than any other $\mu_i$. Thus, the way to think about it is the probability of at least one of the $c$ points being in the top $1/C$, times the probability of such point being $x_0$; plus the probability of at least two points being in the top $1/C$ times the probability of $x_0$ being the second best of the $c$ points sampled... Thus calling $P_i$ the value of \eqref{Probability mu0 > mu1} when taken for points $x_0$ and $x_i,$

\begin{equation}
    \begin{split}
       1- p &\leq \sum_{n=1}^c \binom{c}{n}
        \left(\frac{1}{C}\right)^n\left(1-\frac{1}{C}\right)^{c-n} \cdot\\
        &\cdot \sum_{j_1<...<j_{n-1}} \prod_{k=1}^n (1-P_{j_k})\prod_{i\neq j_1,...,j_{n-1}} P_i.
    \end{split}
\end{equation}

The idea here would be to calculate $c, m_0,...,m_c$ fulfilling the previous equation while minimising $\sum_{i=0}^c m_i$. However, optimizing $m_1,...m_c$ over the previous set can be complicated. This suggests using a different approach, that we review in the following section.

\subsection{\label{sec:Threshold strategy} Threshold strategy}

In the previous section we followed a greedy strategy that was quite efficient when the function $s$ was deterministic, but became more involved when the output of $s$ followed a distribution with variance $\sigma^2$ relatively big compared to the distance $(\mu_0-\mu_1)^2$.

Thus, in this last case, another good strategy could be
\begin{enumerate}
    \item Repeatedly sample uniformly at random points $x\in \mathcal{S}$, and apply $s(x)$, in order to calculate the percentile $1-1/C$ of the evaluation of $s(x)$, $p_C$ with error $\epsilon$ and exponentially good probability $1-p = 1- e^{-\beta}$.
    \item Find a single point $x$ and evaluate $s(x)$ $m_x$ times to make sure that, with exponentially high probability $x$ is in the top $1-1/C$ quantile.
\end{enumerate}

Thus the first step is estimating a percentile. If we sample $n$ points $x_1,...,x_{n}$ and evaluate $s(x_i)$ for each of them, such that $s(x_1)\leq...\leq s(x_{n})$. Then the best estimation of the value in the $1-1/C$ percentile is the weighted average between $x_{\lceil (n+1)(1-1/C) \rceil}$ and $x_{\lfloor (n+1)(1-1/C) \rfloor}$, with weights $|\lceil (n+1)(1-1/C) \rceil - (n+1)(1-1/C)|$ and $|\lfloor (n+1)(1-1/C) \rfloor - (n+1)(1-1/C)|$ respectively. To clarify, suppose for example that we sample 3 points $s(x_1)< s(x_2) <s(x_3)$ from a uniform distribution and we want to calculate the $1/3$ quantile. As the previous calculation and our intuition indicate, it should be the (weighted) average of $x_1$ and $x_2$, because that leaves $1/3$ of the points below it. 

More generally, subindex $i$ follows, for large $n$, a probability distribution $\mathcal{D}$  of $p_C$ being $x_i$. Such probability distribution is a normal distribution with average $\mu = n p_C$ and variance $\sigma^2 = n p_C (1-p_C) $. In other words, the variable 
\begin{equation}
    z = \frac{i -n p_C}{\sqrt{np_C(1-p_C)}}
\end{equation}
follows the normal distribution $\mathcal{N}(\mu=0,\sigma^2=1)$. Since we want to estimate $p_C$ with error $\epsilon' = \epsilon/C$ and exponentially high probability $1-p$, this means
\begin{equation}
\begin{split}
    P\left(\frac{(1-\epsilon')p_C n-p_C n}{\sqrt{np_C(1-p_C)}}\leq z \leq \frac{(1+\epsilon')p_C n-p_C n}{\sqrt{np_C(1-p_C)}}\right)\\
    \geq 1-p
\end{split}
\end{equation}
The reason for choosing $\epsilon' = \epsilon/C$ is to be able to compare with the quantum algorithm. Simplifying,
\begin{equation}
    P\left(-\epsilon' \sqrt{n}\sqrt{\frac{p_C }{1-p_C}}\leq z \leq \epsilon' \sqrt{n}\sqrt{\frac{p_C }{1-p_C}}\right)
    \geq 1-p.
\end{equation}
Calculating the left-hand side of the equation, as $\mu_z=0$ and $\sigma_z = 1$
\begin{equation}
    \begin{split}
        &\left[\frac{1}{2}+\frac{1}{2}erf\left(\frac{z-\mu_z}{\sqrt{2}\sigma_z}\right)\right]_{-\epsilon' \sqrt{n}\sqrt{\frac{p_C }{1-p_C}}}^{+\epsilon' \sqrt{n}\sqrt{\frac{p_C }{1-p_C}}}=\\
        &\frac{1}{2}\left[erf\left(\frac{\epsilon'\sqrt{np_C}}{\sqrt{2(1-p_C)}}\right)-erf\left(\frac{-\epsilon'\sqrt{np_C}}{\sqrt{2(1-p_C)}}\right)\right]=\\
        & erf\left(\frac{\epsilon'\sqrt{np_C}}{\sqrt{2(1-p_C)}}\right) \geq 1- e^{-\beta}.
    \end{split}
\end{equation}
the last equal due to the error function being odd.
This means we need the number of samples to grow like
\begin{equation}
    n \geq  2\frac{1-p_C}{p_C}\frac{1}{\epsilon'^2}\left(erf^{-1}(1-e^{-\beta})\right)^2,
\end{equation}
where $erf^{-1}$ is the inverse error function, and $(1-p_C)/p_C = (C-1)^{-1}$. 

As
\begin{equation}
    \lim_{\beta \rightarrow \infty} \frac{erf^{-1}(1-e^{-\beta})}{\sqrt{\beta}}=1,
\end{equation}
and $\epsilon'=\epsilon/C $, that means that $n = O(\epsilon^{-2}\beta C)$. As a second step we need to calculate the number of times we have to call the sample function $s$ over a candidate in order to certify that with exponentially high probability it is above the percentile $1-1/C$. Using the central limit theorem again, the $m$-sample average $\hat{\mu}_x$ follows a normal distribution $\mathcal{N}(\mu_x,\sigma^2/m)$. So, given a threshold $T$ which indicates the upper part of the confidence interval of the percentile $1-1/C$; and supposing $\mu_x > T$, we want 
\begin{equation}
    \int_T^\infty \mathcal{N}(\mu_x,\sigma^2/m) \geq 1- p = 1- e^{-\beta}.
\end{equation}
The left hand side can be written as
\begin{equation}
\begin{split}
    &\left[\frac{1}{2} + \frac{1}{2}erf\left(\frac{z-\mu_x}{\sqrt{2m}\sigma}\right)\right]_{T}^{\infty}=\\
    &\frac{1}{2}-\frac{1}{2}erf\left(\frac{T-\mu_x}{\sqrt{2m}\sigma}\right).
\end{split}
\end{equation}
From this it can be calculated that 
\begin{equation}
    m \geq \frac{2\sigma^2}{(\mu_x-T)^2}(erf^{-1}(1-2e^{-\beta}))^{-2},
\end{equation}
and as
\begin{equation}
    \lim_{\beta \rightarrow \infty} \frac{erf^{-1}(1-2e^{-\beta})}{\sqrt{\beta}}=1,
\end{equation}
we have $m = O(\beta)$. Over how many points $x$ do we have to repeat this procedure until we make sure that $x$ is over the threshold? As described just after \eqref{complexity C}, one would need to repeat this over $O(C)$ points, although in practice it may be much less given the information we have of the previous steps. This step then has complexity $O(\beta C)$.
Thus the overall complexity of this procedure is $O(\epsilon^{-2}\beta C)$.

\subsection{\label{sec:Quantum_threshold_strategy}Quantum Threshold Strategy}

Can we do better if we use quantum methods? The answer is that we can get a quadratic advantage over some parameters: we will have an overall complexity of $O(\epsilon^{-1}C\beta)$. 

Before explaining how to do it, we need to define the equivalent version of the probabilistic oracle $s$. In this case $s: \ket{x}\rightarrow \sum_{s(x)} \sqrt{p_{x,s(x)}}\ket{x}\ket{s(x)} $, where $p_{x,s(x)}$ are the probabilities and follow the distribution $\mathcal{D}(\mu_x,\sigma^2)$.

The required steps are the following

\begin{enumerate}
    \item Use bisection search, amplitude estimation \cite{Amplitude_estimation} and the median lemma from \cite{nagaj2009fast} to estimate, with success probability $1-p= 1-2^{-\beta}$ and error $\epsilon$ a threshold $T$ for the percentile $1-1/C$ of the Image of $s$.
    \item Use Amplitude Estimation \cite{Amplitude_estimation} to mark any $\ket{x}$ for which the amplitude over threshold $T + \epsilon$ is greater than $1/\sqrt{2}+\epsilon$ . Use and the median lemma from \cite{nagaj2009fast} to achieve an exponential high success probability $1-2^{-\beta}$. The time complexity would be $O(\beta \epsilon)$. 
    \item Use Amplitude Amplification \cite{Amplitude_estimation} to find all such points in time $O(\sqrt{C})$.
\end{enumerate}

Since the median lemma is not widely known, let us state it here without proof.
\begin{Lemma}
\textbf{(Median Lemma)} \cite{nagaj2009fast} Consider a sequence $\{\phi'_k\}$ for $k=1...\beta$. Let the probability that $\phi'_k$ does not belong to $(\phi_L, \phi_R)$ be smaller $\delta < 1/2$. Then, the probability that the median of $\{\phi'_k\}$ falls out of $(\phi_L, \phi_R)$ can be bounded above by 
\begin{equation}
    p_{fail}=\frac{1}{2}\left(2\sqrt{\delta(1-\delta)}\right)^{-\beta}\leq 2^{-\beta-1}.
\end{equation}
\qed
\end{Lemma}

This lemma can be used to reduce, in linear time $O(\beta)$, the probability that Amplitude Estimation with error $\epsilon$ fails. On the other hand, the complexity of Amplitude Estimation is $O(\epsilon^{-1})$, for a given error $\epsilon$.

This means that given an error tolerance $\epsilon/C$, and a probability of failure $2^{-\beta}$ we can perform Amplitude Estimation with complexity $O(\epsilon^{-1} \beta C)$. Finally, to fully understand the step $1$ above, we need to explain what we mean by bisection search. The idea is to propose a trial threshold $T'$ and check whether the percentage of points for which $s(x)$ is above $T'$ is greater or smaller than $1/C$. With that one may refine the initial guess of $T'$ until we know that the percentage of points for which $s(x)>T$ is between $(1-\epsilon)/C$ and $(1+\epsilon)/C$ with exponentially high probability $1-e^{-\beta}$. The cost of bisection search is well known to be logarithmic in the precision.

Overall we can see the complexity of each strategy in the table \ref{Tabla_complexities}.

\begin{table}
 \begin{tabular}{|c | c | c |} 
 \hline
  Complexity & Deterministic & Non-deterministic \\ 
 \hline
 Classical Greedy & $O(\beta C)$ & $\Omega(\beta^2 C)$ \\ 
 \hline
  Classical Threshold & $O(\epsilon^{-2}\beta C)$ & $O(\epsilon^{-2}\beta C)$ \\
 \hline
 Quantum Threshold & $O(\epsilon^{-1}\beta C)$ & $O(\epsilon^{-1}\beta C)$ \\ 
 \hline
\end{tabular}
 \caption{ \label{Tabla_complexities}Complexities of different algorithms. Notice that the step of determining the threshold is dominating in the second and third strategies. However, there is also a difference in the step of determining a point over the threshold. Classically, the complexity would be $O(\beta C)$, whereas quantumly it would be $O(\epsilon^{-1}\beta \sqrt{C})$.}
\end{table}

\section{\label{sec:Norm_appendix}The norm of a vector after quantum linear algebra techniques}

In the appendix A of \cite{FEM} it is explained how to estimate the norm of the solution vector of the HHL algorithm, as we already indicated in the equation \eqref{norm solution HHL}. However, at the end of the appendix they indicated that they were not sure how to estimate the norm of the solution using any other improvements to HHL such as \cite{Ambainis, QLSAchilds}. Here we explain how to estimate the norm of any vector $M\ket{v}$, for $M$ any matrix that we can decompose as a linear combination of unitaries $M= \sum_i \alpha_i U_i$, and $\ket{v}$ a pure quantum state; or when we apply the Chebyshev approach indicated in the main text.

Firstly, let us briefly review the simplest case of LCU technique. Let $M= U_0 +U_1$. Then we can perform
\begin{equation}
    \ket{0}\ket{v}\xrightarrow{H} \frac{1}{\sqrt{2}}(\ket{0}+\ket{1})\ket{v}. \label{applying V for two}
\end{equation}
Now, we perform the so called multi-$U$ operator $\sum_i \ket{i}\bra{i}\otimes U_i$,
\begin{equation}
    \frac{1}{\sqrt{2}}(\ket{0}U_0+\ket{1}U_1)\ket{v},
\end{equation}
And performing a second Hadamard over the first register we get
\begin{equation}
    \frac{1}{2}\left[\ket{0}(U_0+U_1)+\ket{1}(U_0-U_1)\right]\ket{v}.
\end{equation}
Postselecting on measuring $\ket{0}$ on the first register we get a state proportional to $M\ket{v}$. To calculate the probability, let us write it in density matrix form
\begin{equation}
\begin{split}
    \rho = \frac{1}{4}\left[\ket{0}\bra{0}(U_0+U_1)\ket{v}\bra{v}(U_0+U_1)^\dagger\right.\\
    +\ket{0}\bra{1}(U_0+U_1)\ket{v}\bra{v}(U_0-U_1)^\dagger\\
    +\ket{1}\bra{0}(U_0-U_1)\ket{v}\bra{v}(U_0+U_1)^\dagger\\
    \left. +\ket{1}\bra{1}(U_0-U_1)\ket{v}\bra{v}(U_0-U_1)^\dagger \right].
\end{split}
\end{equation}
To calculate the probability of measuring 0, we need to measure the norm of $P_0\rho P_0$, where $P_0 = \ket{0}\bra{0}\otimes \mathbf{1}$, taking the trace, which happens to be $\braket{v|(U_0+U_1)^2|v}/4$. The probability of measuring a $\ket{0}$ is then $\braket{v|(U_0+U_1)^2|v}/4$. This same calculation is performed to calculate the probability of measuring a $\ket{0}$ in the Swap Test. 

Let us use this to generalise to the setting when $M= \sum_i \alpha_i U_i$. The first step, analogous to \eqref{applying V for two} is applying an operator $V$ that prepares the coefficients
\begin{equation}
    \ket{0}\ket{v}\xrightarrow{V\otimes \mathbf{1}} \frac{1}{\sqrt{\sum_i \alpha_i^2}}\sum_i \alpha_i \ket{i}\ket{v} .
\end{equation}
This implies that we have to correct the norm we will measure with $\sqrt{\sum_i \alpha_i^2}$. Then, we apply the multi-$U$, $\sum_i \ket{i}\bra{i} \otimes U_i$.
\begin{equation}
    \frac{1}{\sqrt{\sum_i \alpha_i^2}}\sum_i \alpha_i \ket{i}U_i \ket{v}.
\end{equation}
Next we perform a Hadamard gate, $H\ket{i}= \sum_j (-1)^{i\cdot j}\ket{j}$ over the first register, 
\begin{equation}
    \frac{1}{\sqrt{\sum_i \alpha_i^2}}\sum_{i,j} (-1)^{i\cdot j} \alpha_i \ket{j}U_i \ket{v}.
\end{equation}
Using the same argument as we did in the simpler case, the amplitude of the first register being in state $\ket{0}$ is 
\begin{equation}
    A_0 = \frac{1}{\sqrt{\sum_i \alpha_i^2}} \left|\left|\left(\sum_{i} \alpha_i U_i\right) \ket{v}\right|\right|.
\end{equation}
Therefore, in the same way that the norm of the solution can be calculated using \eqref{norm solution HHL}, we can calculate the norm of the solution in this case using 
\begin{equation}
    ||M v|| =  A_0 ||v|| \sqrt{\sum_i \alpha_i^2} . \label{norm linear combination of unitaries}
\end{equation}
Clearly, estimating a given amplitude requires the use of Amplitude Estimation procedure, with cost $O(\epsilon^{-1})$, and can be applied to the procedure of the Quantum Linear System Algorithm of \cite{QLSAchilds} for example.

Next we want to calculate the factor by which to correct the norm of a vector $\ket{v}$ after applying a Chebyshev polynomial according to \eqref{Quantum walk}. First notice that if $\ket{v} = \ket{\lambda}$, for $\lambda = 1$ the largest eigenvalue of $M$, then using \eqref{Quantum walk} will result in staying the same state $\ket{\lambda}$, due to Lemma \ref{Lemma 15}. This implies that we will have to correct the norm of the solution by the largest eigenvalue $\lambda_{max}$ that will divide $M$ in order to ensure that the largest eigenvalue in $M/\lambda_{\max}$ is 1.

In such case, one can calculate the equivalent of \eqref{norm linear combination of unitaries} as 
\begin{equation}
    ||M v|| =  A_0 ||v|| \lambda_{\max}, 
    \label{norm Chebyshev}
\end{equation}
where $A_0$ is the estimated amplitude of the correct state in \eqref{Quantum walk}.

In general though, $\lambda_{\max}$ is not known. In such case, \cite{QLSAchilds} proposes making all entries in the matrix $M$ smaller than $1/d$, $d$ the sparsity. Then, they argue, the maximum eigenvalue will be in the interval $(-1,1)$ and Lemma \ref{Lemma 15} can still be used. In such case, we will use \eqref{norm Chebyshev} but substituting $\lambda_{\max}$ by the factor that makes all entries smaller than $1/d$. Notice also that when estimating the norm of the solution of the HHL algorithm, we supposed that the maximum eigenvalue was no larger than 1. Therefore, a similar treatment to what we do here should be also applied there.

\section{\label{sec:dequantization}Dequantization of our algorithm}

In order to explain how to reduce the initially exponentially quantum advantage of our algorithm to a quadratic one, we must define the concept of dequantization, first explained in reference \cite{ewin-tang}. A perhaps more pedagogical introduction on the topic might be found in \cite{ewin-tang-pedagogical}, from where the definitions and tools presented are taken from.

The first thing we have to do is to define how to define \textit{query access} and \textit{query and sample access}.
\begin{Definition}
Given a vector $\vec{x}\in \mathbb{C}^n$ for large $n$ we say that we have \textit{query access}, denoted $Q(\vec{x})$ when on input $i$ we can efficiently compute $x_i$.

For that vector, we may say we have \textit{sample and query access}, denoted by $SQ(\vec{x})$ when:
\begin{enumerate}
    \item We have query access to $\vec{x}$: given $i$ we can efficiently compute $x_i$.
    \item We can produce random samples $i\in [n]$ with probability $|x_i|^2/||\vec{x}||^2$.
    \item We can query for $||\vec{x}||$. If we can only query for an upper bound $(1+\nu)||\vec{x}||$, we denote it by $SQ^\nu (\vec{x})$.
\end{enumerate}
For a matrix $A$ we say we have sample and query access when we have such for each row (as vectors) and for the vector made of the norms of the rows.
\end{Definition}
In the first appendix of \cite{tang2018quantum} it is explained how having query and sample access should be seen as the classical equivalent of quick state preparation using several methods such as quantum RAM. Since this is the method we use in our algorithm, sample and query access to the inputs should also be assumed.

This is enough to define what we mean by dequantize an algorithm
\begin{Definition}
Given a quantum algorithm $\mathcal{A}$ with $O(T)$-time preparation input $\ket{\phi_1},...\ket{\phi_c}$ and output a state $\ket{\psi}$ or value $\lambda$. We say that we can dequantize it when there is a classical algorithm $A$ that, given $O(T)$-time access $SQ(\phi_1)$...$SQ(\phi_c)$, can output $SQ^\nu(\psi)$ or $\lambda$ with similar guarantees and polynomial slowdown.
\end{Definition}

The dequantization model rests on three linear algebra techniques. The first one is a dequantization of the swap test:
\begin{Lemma} (Inner product) Proposition 4.2 in \cite{ewin-tang}.

Given $SQ^\nu(\vec{x})$ and $Q(\vec{y})$ we can estimate $\braket{\vec{x}|\vec{y}}$ to precision $\epsilon$ and probability $\geq 1-\delta$ in time $O\left(T\epsilon^{-2}\log \delta^{-1}\right)$.
\end{Lemma}
The corresponding algorithm is indicated in Algorithm 1 in \cite{tang2018quantum}.

The second is a way to estimate the inner product between a matrix and a vector. It relies on using the same technique as the previous Lemma, but noticing that $\ket{Vw}= \bra{V}(\ket{w}\otimes \bm{1})$.
\begin{Lemma} (Thin matrix-vector product) Proposition 4.3 in \cite{ewin-tang}.

Given a matrix $V\in \mathbb{C}^{n\times k}$, $w\in \mathbb{C}^k$, and given $SQ(V^\dagger)$ and $Q(w)$, we can obtain $SQ^\nu(Vw)$ with success probability $\geq 1-\delta$ and complexities
\begin{enumerate}
    \item Query in time $O(Tk)$.
    \item Sample in time $O(Tk^2 C(V,w) \log \delta^{-1})$.
    \item Query the norm in time $O(Tk^2 C(V,w) \nu^{-2}\log \delta^{-1})$.
\end{enumerate}
$C(V,w) = \sum ||w_i V_{*,i}||^2/|Ww|^2$, and $V_{*,i}$ is the $i$-th column of $V$.
\end{Lemma}

Finally, a modified version of the Frieze, Kannan, and Vempala algorithm, that allows to obtain a low rank approximation of a low rank matrix:

\begin{Lemma} (Low rank approximation) Theorem 4.4 in \cite{ewin-tang}.

Suppose $O(T)$-time $SQ(A)$, $A\in \mathbb{C}^{n\times d}$; a singular value threshold $\sigma$ and an error parameter $\epsilon \in (0, \sqrt{\sigma/||A||_F}/4]$. Let $K = ||A||_F^2/\sigma^2$. Then, in time
\begin{equation}
    O\left(\frac{K^{12}}{\epsilon^6}\log^3\delta^{-1} + T \frac{K^8}{\epsilon^4}\log^2\delta\right)
\end{equation}
we output $SQ(S)$, $S\in \mathbb{C}^{q\times n}$, $U\in \mathbb{C}^{q\times l}$, $\Sigma\in \mathbb{R}^{l\times l}$, with $l = \Theta(K^4\epsilon^{-2}\log^2 \delta^{-1})$. These matrices implicitely describe the low rank approximation of $A$, $D = A V V^\dagger$, with $V = S^\dagger U \Sigma^{-1}$. Additionally, with probability $\geq 1-\delta$, $||A-D||^2_F\leq||A-A_l||^2_F+\epsilon||A||_F^2$.
\end{Lemma}

Using the three previous techniques, we can dequantize our algorithm. Clearly, to calculate $P_c(\vec{x}_{n+1})$ we only require the first Lemma above, dequantizing the swap test.

Second, using the techniques developed in \cite{chia2018quantum,arrazola2019quantum} we can dequantize the solution of the linear system of equations \eqref{system of equations}. This requires the use of the Low rank decomposition and the inner product. Using the low rank decomposition we decompose the matrix $F$ and obtain $SQ(V)$. Then
\begin{equation}
    F^{-1}\vec{y} = (F^T F)^{-1} F^T = \sum_i \frac{1}{\Sigma_{ii}}\vec{v}_i \vec{v}_i^T F^T \vec{y}
\end{equation}
We may then estimate $\vec{v}_i^T F^T \vec{y} = Tr (F^T \vec{y}\vec{v}_i^T)$. Finally, using the inner product Lemma, we may estimate the inner product between $F^T = \sum_{jk} F_{kj}\ket{jk}$ and $\vec{y}\cdot\vec{v}_i^T = \sum_{jk} y_j (v_i)_k\ket{jk}$.

Finally we have to apply matrix $A$ from \eqref{new Matrix system} over the vector resulting from the previous paragraph, and calculate its norm. This may be done in an analogous way to how the Quantum Nearest Centroid is dequantized in \cite{tang2018quantum}. Call $w = \ket{b,\vec{\alpha}}$, then
\begin{equation}
    ||A\ket{b,\vec{\alpha}}||^2 = \braket{b,\vec{\alpha}|A^\dagger A|b,\vec{\alpha}} = \braket{a|b}
\end{equation}
for
\begin{equation}
    a = \sum_{i}\sum_j \sum_k A_{ji}||A_{k,*}||\ket{i}\ket{j}\ket{k},
\end{equation}
\begin{equation}
    b = \sum_{i}\sum_j \sum_k \frac{w_j w_k  A_{ki}}{||A_{k,*}||}\ket{i}\ket{j}\ket{k}.
\end{equation}

So, clearly it is easy dequantize our algorithm. The only question that remains is the complexity. The most expensive part of the algorithm is the solution of the linear system of equations, with complexity $O(||F||^6 k^6 \kappa^{16} \epsilon^{-6})$, $k$ the rank. To the previous term, we may add an additional complexity of $O(\epsilon^{-2})$ due to the inner product estimate. Now it becomes apparent that the quantum algorithm has a polynomial speedup when compared against its dequantized counterpart. 
Notice that here $k$ taken constant since we have only taken care of $O(1)$ eigenvalues of size $O(1)$ in the original matrix. The factor $\kappa_{\text{eff}}^{-1} = O(1)$ might be compared with the $\sigma$ factor in the previous low rank approximation lemma.  This concludes the dequantization of our algorithm.
\end{document}